\documentclass[twocolumn, english, aps, prl, superscriptaddress]{revtex4-1}
\usepackage[T1]{fontenc}
\usepackage[latin9]{inputenc}
\usepackage{babel}
\usepackage{amsmath}
\usepackage{amssymb}
\usepackage{graphicx}
\usepackage{esint}
\usepackage[unicode=true,pdfusetitle,
 bookmarks=true,bookmarksnumbered=false,bookmarksopen=false,
 breaklinks=false,pdfborder={0 0 1},backref=false,colorlinks=false]
 {hyperref}
\usepackage{wasysym}
\usepackage{siunitx}
\usepackage{babel}
\usepackage[caption=false]{subfig}

\providecommand{\tabularnewline}{\\}


\begin{document}

\title{Cavity optomagnonics with magnetic textures: coupling a magnetic vortex to light}

\author{Jasmin Graf}

\affiliation{Max Planck Institute for the Science of Light, Staudtstraße 2, 91058 Erlangen, Germany}

\author{Hannes Pfeifer}

\affiliation{Max Planck Institute for the Science of Light, Staudtstraße 2, 91058 Erlangen, Germany}

\author{Florian Marquardt}

\affiliation{Max Planck Institute for the Science of Light, Staudtstraße 2, 91058 Erlangen, Germany}

\affiliation{Institute for Theoretical Physics, University Erlangen-Nürnberg, Staudtstraße 2, 91058 Erlangen, Germany}

\author{Silvia {Viola Kusminskiy}}

\affiliation{Max Planck Institute for the Science of Light, Staudtstraße 2, 91058 Erlangen, Germany}

\begin{abstract}
Optomagnonic systems, where light couples coherently to collective excitations in magnetically ordered solids, are currently of high interest due to their potential for quantum information processing platforms at the nanoscale. Efforts so far, both at the experimental and theoretical level, have focused on systems with a homogeneous magnetic background. A unique feature in optomagnonics is however the possibility of coupling light to spin excitations on top of magnetic textures. We propose a cavity-optomagnonic system with a non homogeneous magnetic ground state, namely a vortex in a magnetic microdisk. In particular we study the coupling between optical whispering gallery modes to magnon modes localized at the vortex. We show that the optomagnonic coupling has a rich spatial structure and that it can be tuned by an externally applied magnetic field. Our results predict cooperativities at maximum photon density of the order of $\mathcal{C}\approx10^{-2}$ by proper engineering of these structures. 
\end{abstract}

\maketitle

\emph{Introduction.\textendash{}} Optomagnonics is an exciting new field where light couples coherently to elementary excitations in magnetically ordered systems. The origin of this photon-magnon interaction is the Faraday effect, where the magnetization in the sample causes the light's polarization plane to rotate. Conversely, the light exerts a small effective magnetic field on the material's magnetic moments. Shaping the host material into an optical cavity enhances the effective coupling according to the increased number of trapped photons.

Recent seminal experiments have demonstrated this coupling~\cite{haigh_triple-resonant_2016,osada_cavity_2016,zhang_optomagnonic_2016}. In these, an Yttrium Iron Garnet (YIG) sphere serves as the host of the magnetic excitations and, via whispering gallery modes (WGM), as the optical cavity. The optomagnonic coupling manifests itself in transmission sidebands at the magnon frequency. So far, these experiments have probed mostly the homogeneous magnetic mode (Kittel mode) where all spins rotate in phase~\cite{kittel_theory_1948}. Very recently, optomagnonic coupling to other magnetostatic modes~\cite{sharma_light_2017,osada_orbital_2017} has been demonstrated, albeit still on top of a homogeneous background~\cite{osada_brillouin_2017,haigh_selection_2018}.

The Kittel mode, although it is the simplest one to probe and externally tune, is a bulk mode and has a suboptimal overlap with the optical WGMs living near the surface. Another caveat is the state of the art in terms of sample size, which is currently sub-millimetric. This results in modest values for the optomagnonic coupling and motivates the quest for smaller, micron-sized magnetic samples, as well as for engineering the coupling between magnetic and optical modes. Increasing the currently observed values of optomagnonic coupling is an urgent prerequisite for moving on to promising applications such as magnon cooling, coherent state transfer, or efficient wavelength converters \cite{soykal_strong_2010,huebl_high_2013,zhang_strongly_2014,tabuchi_coherent_2015,haigh_dispersive_2015,zhang_superstrong_2016,zhang_cavity_2016,liu_optomagnonics_2016,tabuchi_quantum_2016,lachance-quirion_resolving_2016,bourhill_ultrahigh_2016,maier-flaig_tunable_2017,harder_damping_2017,pantazopoulos_photomagnonic_2017,morris_strong_2017,sharma_optical_2018}. 

In microscale magnetic samples, the competition between the short-range exchange interaction and the boundary-sensitive demagnetization fields can lead to \emph{magnetic textures}, where the magnetic ground state is not homogeneous~\cite{hubert_alex_magnetic_1998,_principles_????}. A well studied case is that of a thin microdisk, where the magnetization swirls in the plane of the disk and forms a magnetic vortex in the center~\cite{shinjo_magnetic_2000,guslienko_magnetic_2008-1}, see Fig.~\ref{Fig1}a. In the vortex core, the spins point out of plane. 
%-----------------------------------------------------------------------------------------------------------------------------------------------------------------------
\begin{figure}
\includegraphics[width=1\columnwidth]{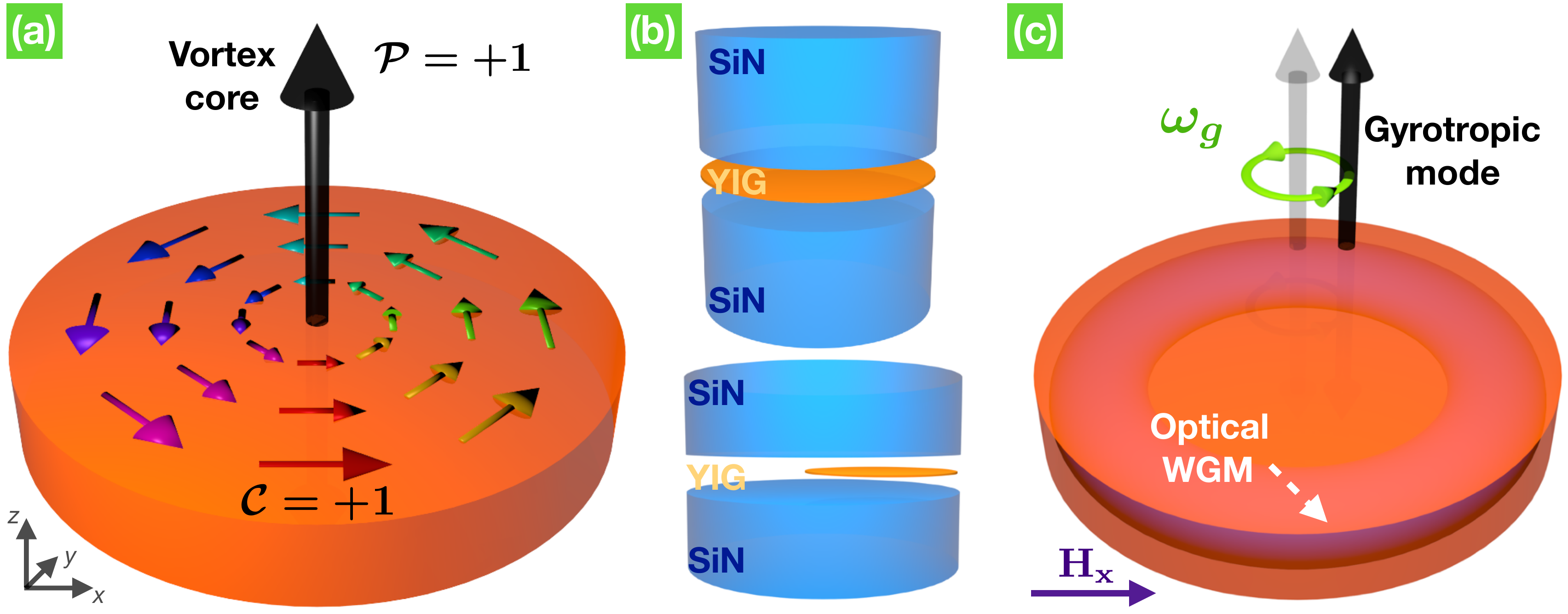} \caption{a) In a magnetic microdisk, the ordered state is a vortex with chirality $\mathcal{C}=\pm1$ and polarity $\mathcal{P}=\pm1$. Envisioned setups: b) A thin YIG disk embedded in an optical cavity of choice. c) A thicker YIG microdisk can support both optical WGM and magnon modes. An external in-plane magnetic field displaces the vortex core in a direction perpendicular to the field.}
\label{Fig1} 
\end{figure}
%-----------------------------------------------------------------------------------------------------------------------------------------------------------------------
Magnetic vortices carry two degrees of freedom: how the magnetization curls (clock or anti-clock wise) defines the \emph{chirality} $\mathcal{C}=\pm1$, while its pointing up or down at the center of the vortex defines the \emph{polarity }$\mathcal{P}=\pm1$. These are robust topological properties and make vortices interesting for information processing~\cite{_principles_????}. Moreover, the position of the vortex can be controlled by an external magnetic field, making this system highly tunable (Fig.~\ref{Fig1}c).

While optomagnonic systems present analogies to optomechanics~\cite{viola_kusminskiy_coupled_2016-1,liu_optomagnonics_2016} (where light couples to phonons~\cite{aspelmeyer_cavity_2014}), the possibility of coupling light to magnetic textures is unique to optomagnonics. In this work we study the optomagnonic coupling in the presence of an inhomogeneous magnetic background in a microdisk geometry (note that this differs from Refs.~\cite{osada_brillouin_2017,osada_orbital_2017,haigh_selection_2018}, where the underlying magnetic ground state is uniform). This is a relevant case to study since: (i) YIG disks at the microscale have been experimentally realized and the presence of magnetic vortices demonstrated~\cite{losby_torque-mixing_2015,collet_generation_2016,zhu_patterned_2017} (ii) a disk supports optical WGMs while reducing the magnetic volume with respect to a sphere, which could lead to larger optomagnonic couplings, and (iii) the spin excitations in the presence of the vortex are qualitatively different from those on top of a homogeneous magnetization. 

We combine analytical methods with micromagnetic and finite-element simulations to derive the spatial dependence and the strength of the optomagnonic coupling. We study two qualitatively different regimes that can be accessed by nanostructure-patterning: a very thin micromagnetic disk embedded in an optical cavity, and a thicker microdisk that also serves as the optical cavity (Fig.~\ref{Fig1}b, c). We demonstrate our method for the coupling between magnon modes localized at the magnetic vortex and the optical WGMs, and predict high values for the optomagnonic coupling and the cooperativity, an important figure of merit in these systems.

\emph{Optomagnonic coupling for magnetic textures.\textendash{}} In a Faraday active material, the electromagnetic energy is modified by the coupling between the electric field and the magnetization~\cite{landau_l._d._electrodynamics_1984}:
\begin{equation}
H_{{\rm MO}}=-i\frac{\theta_{{\rm F}}\lambda_{n}}{2\pi}\frac{\varepsilon_{0}\varepsilon}{2}\int{\rm d}\mathbf{r}~\mathbf{m}(\mathbf{r},t)\cdot[\mathbf{E^{*}}\left(\mathbf{r},t\right)\times\mathbf{E}\left(\mathbf{r},t\right)]\,,\label{eq:UMO-1}
\end{equation}
where $\mathbf{m}(\mathbf{r},t)$ is the local magnetization in units of the saturation magnetization $M_{{\rm s}}$, and we used the complex representation of the electric field, $\left(\mathbf{E^{*}}+\mathbf{E}\right)/2$. The prefactor $\theta_{{\rm F}}\lambda_{n}/2\pi$ ($\sim4\times10^{-5}$ in YIG) gives the Faraday rotation $\theta_{{\rm F}}$ per wavelength $\lambda_{n}$ in the material, $\varepsilon$($\varepsilon_{0}$) is the relative (vacuum) permittivity, and $n=\sqrt{\varepsilon/\varepsilon_{0}}$ the refractive index. Eq.~\eqref{eq:UMO-1} couples the spin density in the magnetic material with the \emph{optical spin density} (OSD), which represents the spin angular momentum density carried by the light field. Quantizing Eq.~\eqref{eq:UMO-1} leads to the optomagnonic Hamiltonian~\cite{viola_kusminskiy_coupled_2016-1}. The coupling is parametric, coupling one local spin operator to two photon operators.

We consider the coupling of the optical fields to spin wave excitations on top of a nonuniform static ground state $\mathbf{m}_{0}(\mathbf{r})$, $\delta\mathbf{m}(\mathbf{r},t)=\mathbf{m}(\mathbf{r},t)-\mathbf{m}_{0}(\mathbf{r})$. For small deviations $|\delta\mathbf{m}|\ll1$ we can express these in terms of harmonic oscillators (magnon modes). Quantizing $\delta\mathbf{m}(\mathbf{r},t)\rightarrow\frac{1}{2}\sum_{\gamma}\left(\delta\mathbf{m}_{\gamma}(\mathbf{r})\hat{b}_{\gamma}e^{-i\omega_{\gamma}t}+\delta\mathbf{m}_{\gamma}^{*}(\mathbf{r})\hat{b}_{\gamma}^{\dagger}e^{i\omega_{\gamma}t}\right)$ and $\mathbf{E}^{(*)}(\mathbf{r},t)\rightarrow\sum_{\beta}\mathbf{E}_{\beta}^{(*)}(\mathbf{r})\hat{a}_{\beta}^{(\dagger)}e^{-(+)i\omega_{\beta}t}$, from Eq. \eqref{eq:UMO-1} we obtain the coupling Hamiltonian $\hat{H}_{MO}=\sum_{\alpha\beta\gamma}G_{\alpha\beta\gamma}^{{\rm }}\hat{a}_{\alpha}^{\dagger}\hat{a}_{\beta}\hat{b}_{\gamma}+{\rm h.c.}$ where $G_{\alpha\beta\gamma}^{{\rm }}=\int{\rm d}\mathbf{r}\,G_{\alpha\beta\gamma}(\mathbf{r)}$ and
\begin{equation}
G_{\alpha\beta\gamma}^{{\rm }}(\mathbf{r)}=-i\frac{\theta_{{\rm F}}\lambda_{n}}{4\pi}\frac{\varepsilon_{0}\varepsilon}{2}\delta\mathbf{m}_{\gamma}(\mathbf{r)}\cdot[\mathbf{E_{\alpha}^{*}}\left(\mathbf{r}\right)\times\mathbf{E}_{\beta}\left(\mathbf{r}\right)]\label{eq:coupling}
\end{equation}
is the local optomagnonic coupling. The Greek subindices indicate the respective magnon and photon modes which are coupled. We use Eq.~\eqref{eq:coupling} to evaluate the coupling between optical WGMs and magnon modes in a YIG microdisk with a magnetic vortex, focusing on magnonic modes localized at the vortex. We study two cases: (i) a thin disk where the problem is essentially 2D, and (ii) a thicker disk, where the \emph{z}-dependence of the problem is non-trivial. The thin disk allows us to compare with analytical approximate results, validating our numerical results.

\emph{Thin Disk.\textendash{}} We consider a magnetic microdisk of thickness $h$ and radius $R$. The characteristic magnetic length scale is the exchange length $l_{{\rm ex}}$ (for YIG: $l_{{\rm ex}}\sim\SI{10}{\nano\meter}$)~\footnote{Another important scale for domain formation is the anisotropy constant$K$, which we assume to be small.}. A vortex is the stable magnetic texture for $h\gtrsim l_{{\rm ex}}$ and $R\gg l_{{\rm ex}}$~\cite{guslienko_magnetic_2008-1}. The lowest excitation mode consists of the vortex's center-of-mass rotating around an axis perpendicular to the disk's plane~\cite{thiele_steady-state_1973,guslienko_eigenfrequencies_2002}, see Fig.~\ref{Fig1}c. The frequency $\omega_{g}$ of this \emph{gyrotropic} mode can be approximated by $\omega_{g}/2\pi\approx\mu_{0}\gamma M_{{\rm s}}\,h/(4\pi R)~$ with $\gamma$ the gyromagnetic ratio~\cite{guslienko_magnetic_2008-1} (for YIG $\omega_{g}/2\pi\approx\SI{1}{\giga\hertz}\cdot h/R$). The excitation is localized at the vortex core, decaying linearly with distance~\cite{SuppA}. 
%-----------------------------------------------------------------------------------------------------------------------------------------------------------------------
\begin{figure}
\begin{centering}
\includegraphics[width=1\columnwidth]{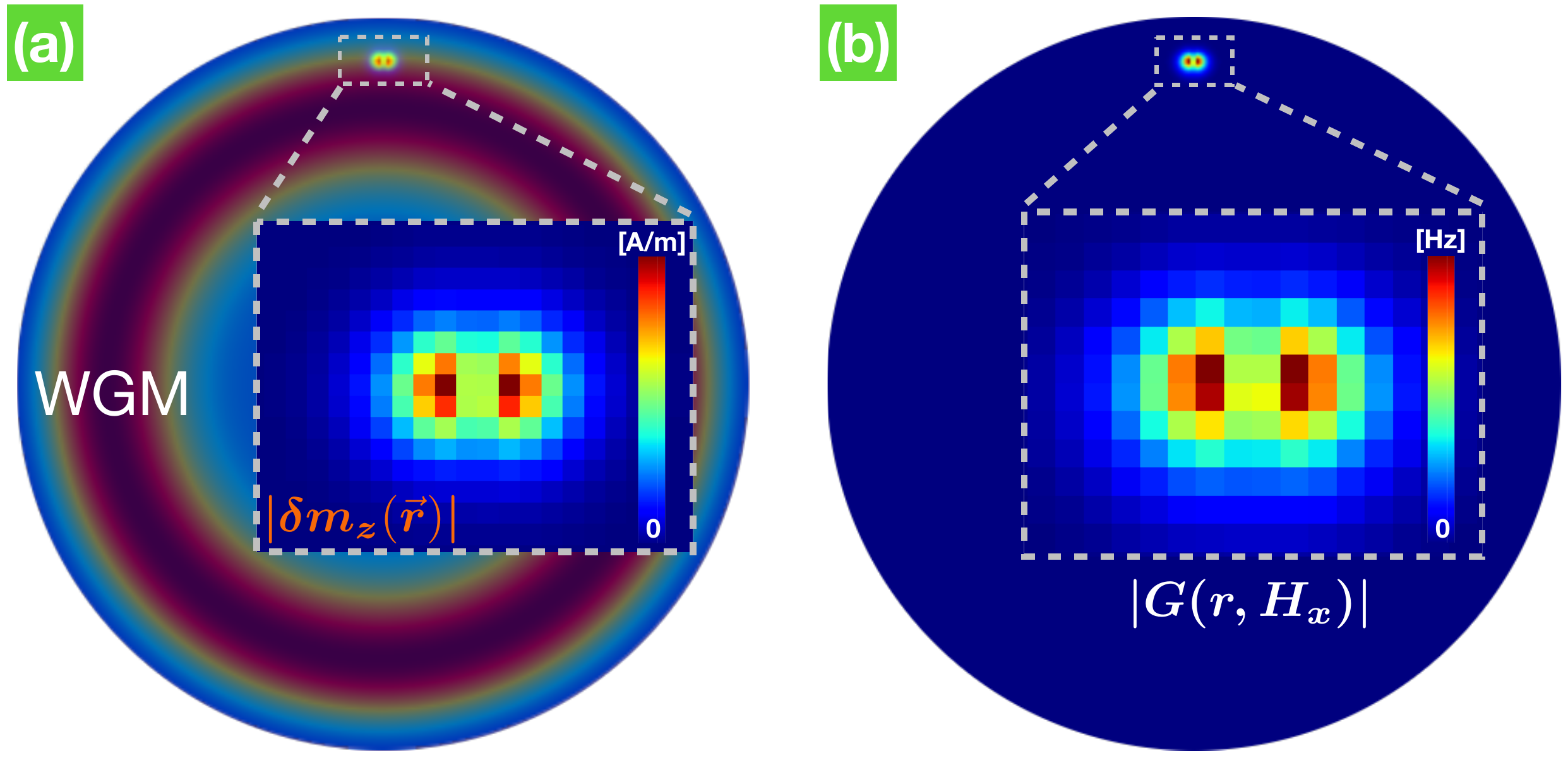}
\par\end{centering}
\caption{Optomagnonic coupling to vortex motion in a thin disk. a) Optical WGM and gyrotropic mode (inset) for a displaced vortex. b) Spatial profile of the optomagnonic coupling. ($R=\SI{1}{\micro\meter}$, $h=\SI{20}{\nano\meter}$, $\omega_{{\rm opt}}/2\pi=\SI{217}{\tera\hertz}$, $\kappa_{{\rm opt}}=\SI{1.51}{\tera\hertz}$, $\omega_{g}/2\pi=\SI{36}{\mega\hertz}$ $H_{x}=\SI{3.3}{\milli\tesla}$).}
 \label{Fig2a}
\end{figure}
%-----------------------------------------------------------------------------------------------------------------------------------------------------------------------
%-----------------------------------------------------------------------------------------------------------------------------------------------------------------------
\begin{figure}
\includegraphics[width=1\columnwidth]{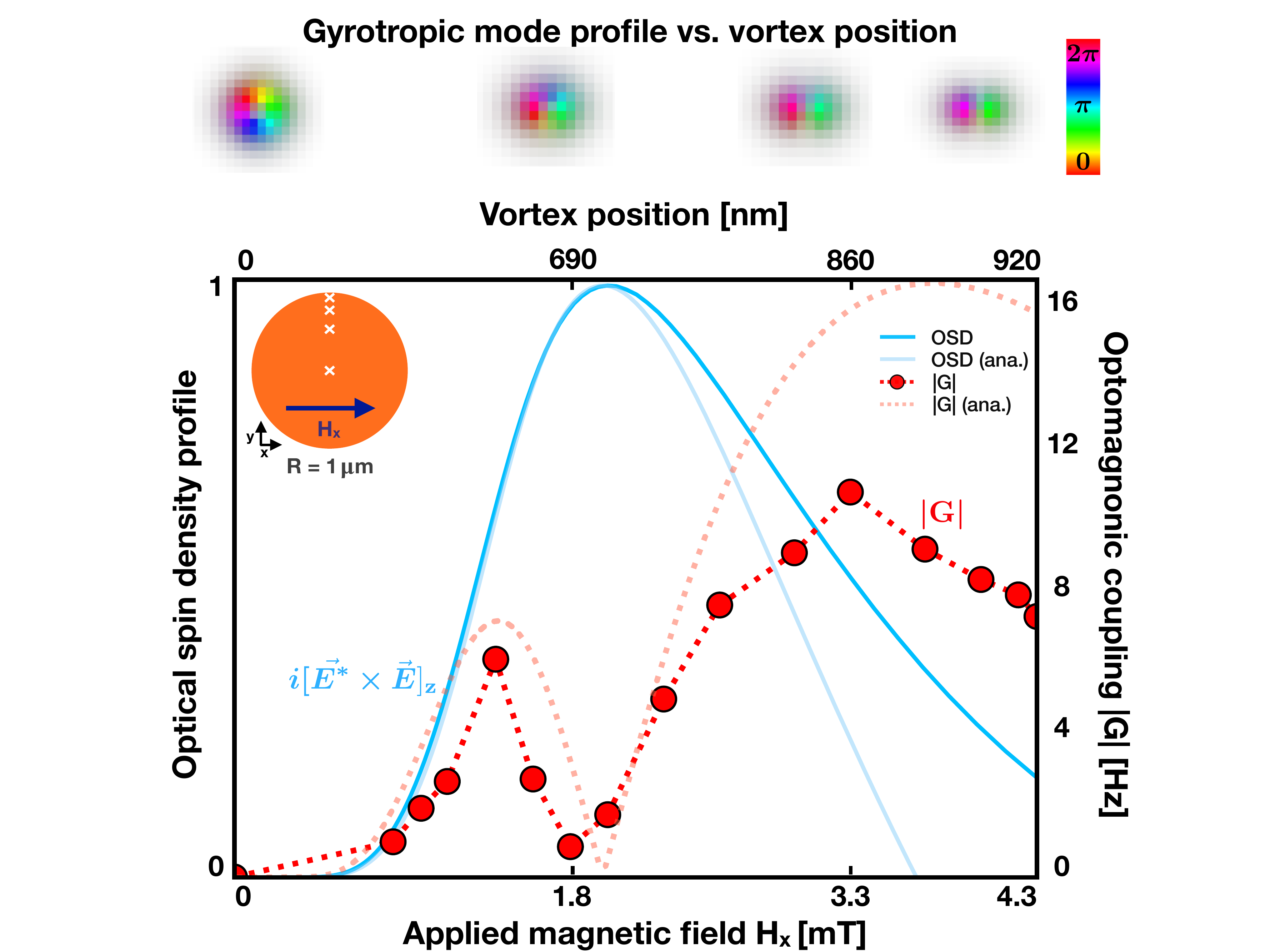}
\caption{Optomagnonic coupling (red dots: numerical, light-red dotted line: analytical) as a function of magnetic field $H_{x}$ (bottom axis, linear scale) or vortex position (top axis) for the thin disk, and OSD (blue solid line: numerical, light blue solid line: analytical) at the position of the vortex, normalized to the maximum OSD. Note the nonlinear dependence of the position of the vortex on $H_{x}$~\cite{SuppE}. The vortex position is shown schematically in the inset. Above the graph, the complex-valued mode function $\delta m_{z}(\mathbf{r})$ shows the distortion of the gyrotropic mode as the vortex moves close to the rim. The analytical model does not consider this nor the decay of the optical field, which accounts for the deviations near the rim of the disk. ($R=\SI{1}{\micro\meter}$, $h=\SI{20}{\nano\meter}$, $\omega_{{\rm opt}}/2\pi=\SI{217}{\tera\hertz}$, $\kappa_{{\rm opt}}=\SI{1.51}{\tera\hertz}$).}
\label{Fig2}
\end{figure}
%-----------------------------------------------------------------------------------------------------------------------------------------------------------------------
The disk also supports optical WGMs. The approximate 2D analytical solution for these is well known~\cite{SuppC}. 
The WGMs can be classified into TM and TE modes, for electric field perpendicular to and in the plane of the disk respectively~\footnote{Note that in some works the opposite convention is used, see e.g.~\cite{osada_cavity_2016}.}. Within this approximation we have two possibilities for finite coupling to the gyrotropic mode: processes involving both TE and TM modes, and those involving only TE modes. For processes involving both TE and TM modes, $\mathbf{E}_{\alpha}^{{\rm TE}*}\times\mathbf{E}_{\beta}^{{\rm TM}}$ lies in the $xy$ plane and therefore can couple to the in-plane component of the gyrotropic mode, which is finite both inside and outside of the vortex core. Processes involving instead only TE modes couple exclusively to the out-of-plane component of the gyrotropic mode, which is finite only inside the vortex core~\cite{SuppA}. 
For a YIG microdisk, the free spectral range $\Delta f_{{\rm FSR}}\approx\SI{0.1}{\tera\hertz}$, which is much larger than the typical gyrotropic frequencies. Therefore, magnon scattering between two energetically distinct optical modes would be allowed either in the sideband unresolved case, or possibly with carefully selected modes of other radial optical quantum numbers. Moreover, using an external magnetic field for frequency-matching can be difficult in these structures, since it would alter the static magnetic texture and consequently the modes. In the following we discuss the case of scattering with one TE mode, which is free from these considerations. This is analogous to single-mode optomechanics~\cite{aspelmeyer_cavity_2014} or optomagnonics~\cite{viola_kusminskiy_coupled_2016-1}, where the system is driven by a laser whose detuning from the optical mode can be made to match the magnon frequency.

Coupling to the gyrotropic mode is only possible if there is an overlap with the WGM. Applying a magnetic field $H_{x}$ along $x$ displaces the vortex up (down) along $y$ for counterclockwise (clockwise) chirality, as the spins try to align with the field. This provides a knob to control the optomagnonic coupling, as we show in the following.
%-----------------------------------------------------------------------------------------------------------------------------------------------------------------------
\begin{figure*}[t!]
\subfloat{\includegraphics[height=0.3\textheight]{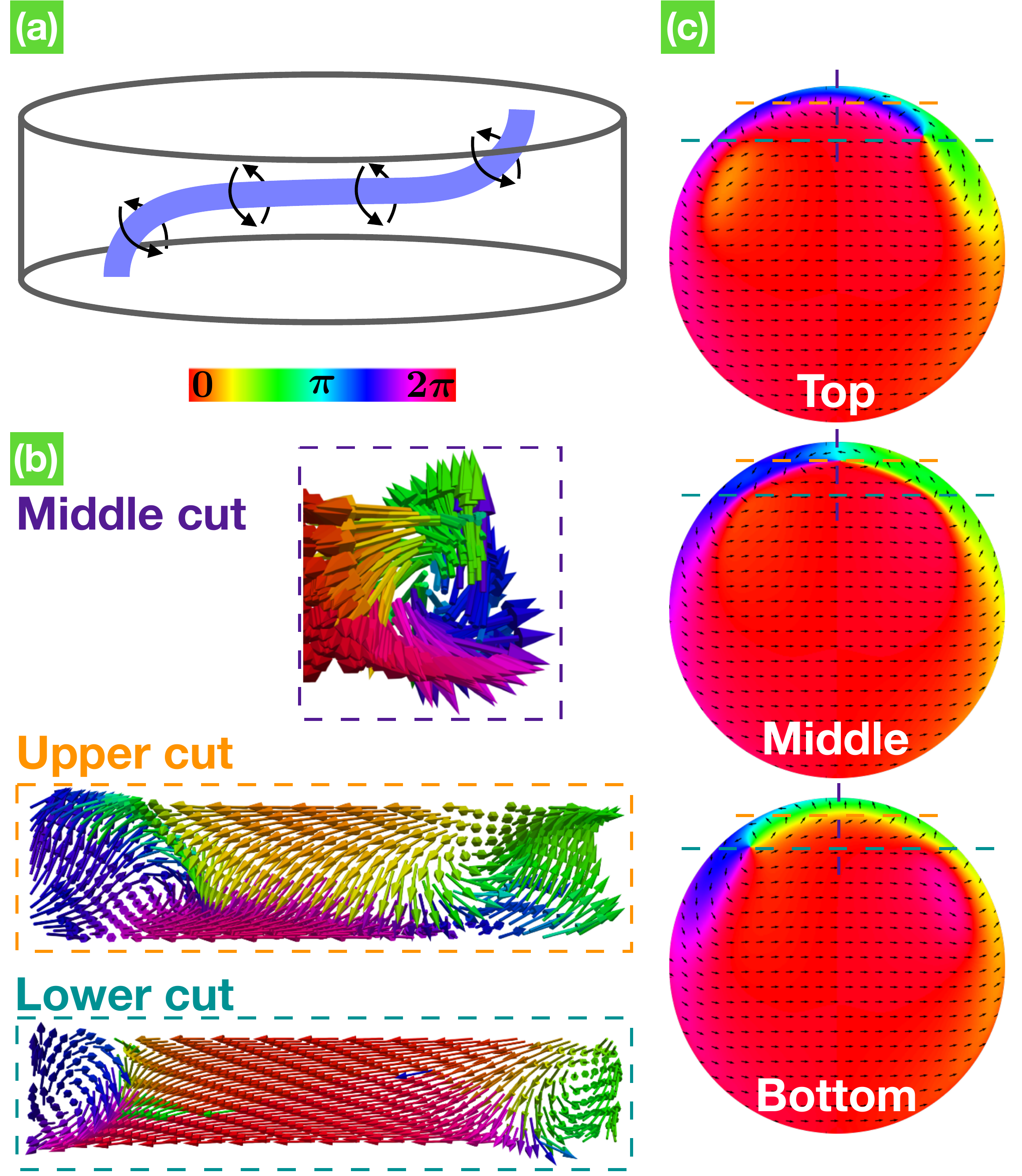}

}\subfloat{\includegraphics[height=0.3\textheight]{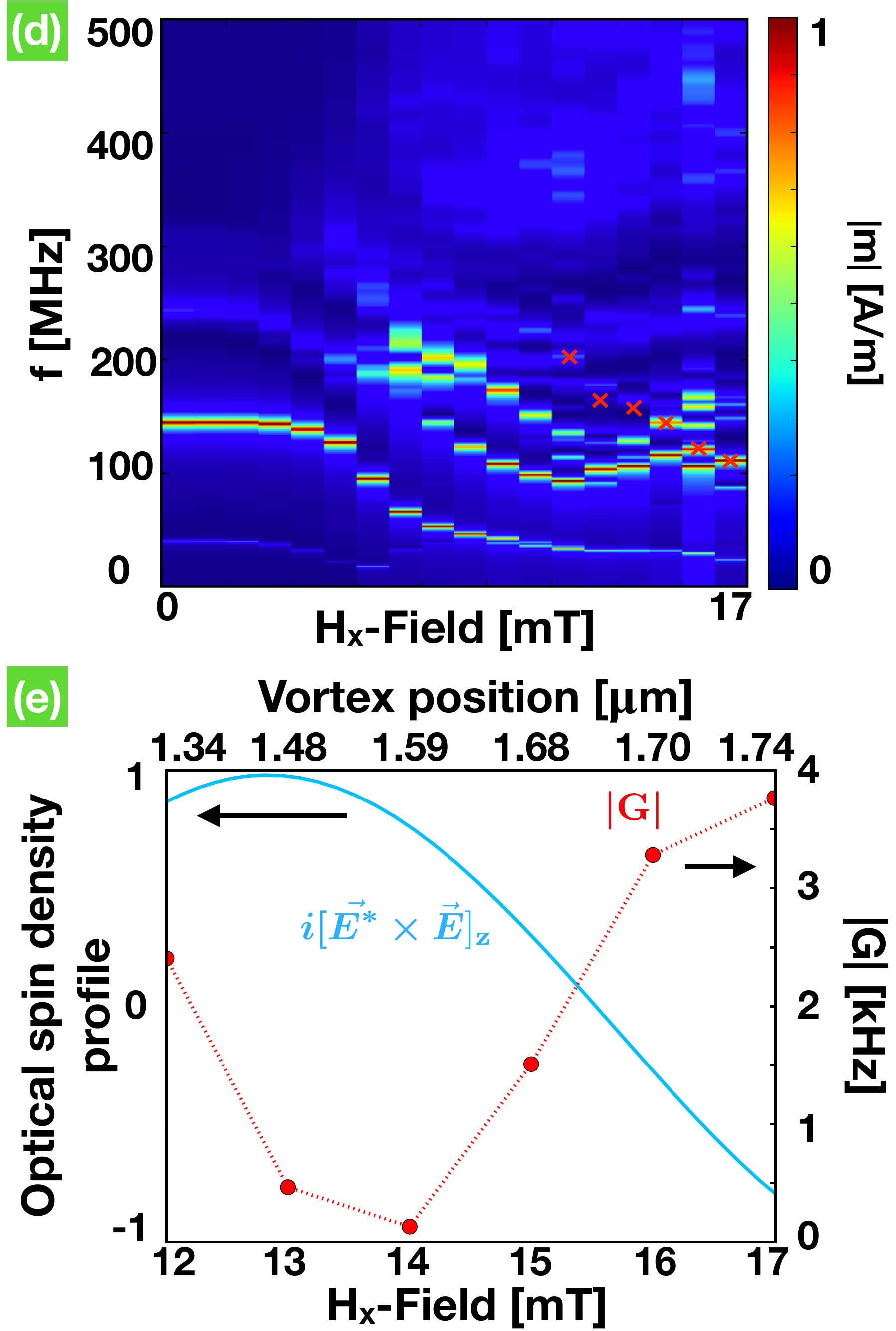}

}\subfloat{\includegraphics[height=0.3\textheight]{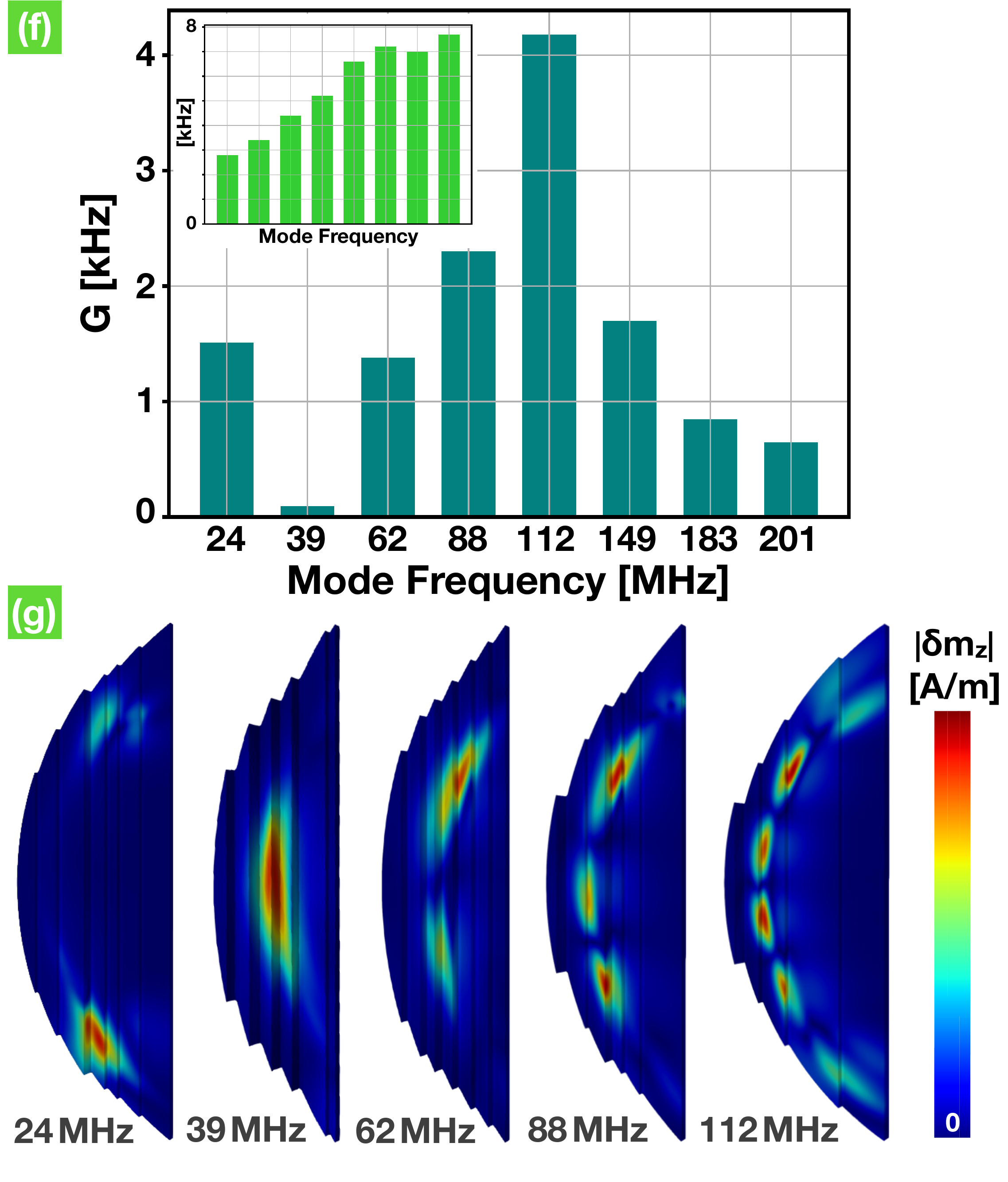}

}\caption{Magnetic background, localized magnon modes, and optomagnonic coupling for a thick disk. a) Sketch of the magnetic vortex. b) and c) Sectional cuts showing the 3D structure of the vortex. d) Magnon spectrum vs. magnetic field. The red crosses indicate the magnon mode considered in plot e). e) Coupling as a function of magnetic field (red dots), and OSD (blue solid line) at the vortex position. f) Coupling as a function of magnon mode frequency. The inset shows $\int_V {\rm d}^3r |G(\mathbf{r})|$ for comparison. g) Profile of the first five excited magnon modes. For b), c), f), and g), $H_{x}=17\,\text{mT}$. All couplings for a TE-WGM of $\omega_{{\rm opt}}/2\pi=\SI{184}{\tera\hertz}$, $\kappa_{{\rm opt}}=\SI{0.03}{\tera\hertz}$. ($R=\SI{2}{\micro\meter}$, $h=\SI{500}{\nano\meter}$).}
\label{Fig3} 
\end{figure*}
%-----------------------------------------------------------------------------------------------------------------------------------------------------------------------
We first note however that a thin YIG microdisk such that $h\gtrsim l_{{\rm ex}}\approx\SI{10}{\nano\metre}$ is a bad optical cavity. To better confine the optical modes, we consider a structure as shown in Fig.~\ref{Fig1}b, such that the YIG disk is placed between two dielectric, non-magnetic disks with same radius and height comparable to $\lambda_{n}$. We chose $\text{Si}_{3}\text{N}_{4}$ (refractive index $n_{{\rm \text{Si}_{3}\text{N}_{4}}}\approx2$) in order to create an almost continuous material for the WGM resonator. Hence the WGMs live in the whole structure, whereas the magnon modes are confined to the thin YIG disk. The gyrotropic mode can overlap with the WGMS for a displaced vortex, see Fig.~\ref{Fig2a}a. We continue to call this mode ''gyrotropic'' since it evolves continuously from the gyrotropic mode at $H_{x}=0$. Whereas its frequency has a light dependence on $H_{x}$, the mode itself is distorted as the rim of the disk is approached. This reflects the deformation of the vortex core into a C-shaped domain wall due to the stronger influence of the demagnetization fields at the nearest boundary~\cite{rivkin_analysis_2007,aliev_spin_2009, SuppD}. 

Fig.~\ref{Fig2a} shows an example of the spatial dependence of the optomagnonic coupling $|G(\mathbf{r},H_{x})|$ for the gyrotropic mode and a WGM. The coupling, given throughout this work as per photon and per magnon, was obtained by combining MuMax$^{3}$~\cite{vansteenkiste_design_2014-1} micromagnetic simulations with finite-element simulations for the optical WGM via Eq.~\eqref{eq:coupling}. Details on the simulations and the normalization procedure are presented~\cite{SuppF, SuppG, SuppH}. 
The total coupling is obtained by integrating over the whole volume. The integration volume is however bounded by the magnon mode volume, $V_{{\rm mag}}$, since it is smaller than the optical mode $V_{{\rm opt}}$. A quick estimate of the maximum coupling is $|G_{{\rm max}}|\approx\sqrt{2g\mu_{B}/(M_{{\rm s}}V_{{\rm mag}})}\left(\theta_{{\rm F}}\lambda_{n}/2\pi\right)\left(V_{{\rm mag}}/V_{{\rm opt}}\right)\hbar\omega_{{\rm opt}}$ ($g$ g-factor, $\mu_B$ Bohr magneton), showing a suppression of the coupling by a factor $\sqrt{V_{{\rm mag}}}/V_{{\rm opt}}$. For the thin disk considered here we find $|G_{{\rm max}}|\approx\SI{30}{\hertz}$, in agreement with the modest maximum value of $|G|\approx\SI{10}{\hertz}$ obtained numerically, see Fig.~\ref{Fig2}. Interestingly, this maximum value is not obtained at the maximum of the OSD, but at points of its maximum slope (as a function of vortex position). This can be understood by noting the antisymmetry under inversion of $\delta m_{z}$ for the gyrotropic mode, which leads to a cancellation when integrated weighted by an isotropic factor. This cancellation is lifted most effectively when the vortex is located at highly anisotropic points of the OSD. Fig.~\ref{Fig2} shows $|G|$ as a function of applied magnetic field, together with the profile for the OSD. This shows clearly that the magnon mode couples effectively to the gradient of the OSD. The value of $|G|$ is therefore completely tunable by an external magnetic field, in contrast to the usual case of magnonic modes on a homogeneous background.

Using the ``rigid vortex'' model~\cite{guslienko_magnetic_2008-1,usov_magnetization_1993} for the magnetics, the optomagnonic coupling for the gyrotropic mode can be obtained analytically~\cite{SuppA, SuppB, SuppD}. Using that the vortex core radius $b$ is small, the first non-zero contribution to the coupling in a Taylor expansion is 
\begin{equation}
\left|G_{m}(s)\right|\approx  10^{-2}\theta_F\lambda_n \sqrt{\frac{g\mu_Bh}{M_{\rm{s}}}} b^2 \varepsilon_0\varepsilon\left|\partial_{s} \left[(\mathbf{E}_{m}^{{\rm TE}*}\times\mathbf{E}_{m}^{{\rm TE}})_z\right]\right|
\end{equation}
with $s$ the vortex position and $m$ the WGM label. This confirms the coupling to the gradient of the OSD. This simplified analytical model is in good agreement with the simulations, see Fig.~\ref{Fig2}. 

\emph{Thick Disk.\textendash{}} The magnetic texture can be considered independent of height when $h$ is only a few $l_{{\rm ex}}$. Increasing the height of the disk leads to more complex magnetic textures and the appearance of magnon flexural modes along the $z$-direction~\cite{ding}, which can hybridize with in-plane modes~\cite{noske_three-dimensional_2016}. Although this effect is already present for $H_{x}=0$, it is even more striking for finite external field. We discuss this regime in the following.

We consider a ``thick'' microdisk such that $h\gg l_{{\rm ex}}$ in an applied external field $H_{x}$. In this case the vortex ``snakes'' from top to bottom of the disk, see Fig.~\ref{Fig3}a-c. This results in highly complex magnon modes, which we obtain by micromagnetic simulations. The spatial structure for the first excited modes is shown in Fig.~\ref{Fig3}g. We interpret these as flexural modes of the vortex core, possibly hybridized with the gyrotropic mode. The optomagnonic coupling for these modes at a fixed $H_{x}$ is presented in Fig.~\ref{Fig3}f. We observe that (i) we obtain values for the coupling in the kHz range, and (ii) the value of the coupling has a non-monotonic dependence on the mode number, due to cancellation effects, as can be seen when compared with the integrated absolute value of the coupling. This system shows also tunability by an external magnetic field, and the coupling is governed by the gradient of the OSD, see Fig.~\ref{Fig3}. Taking the Gilbert damping coefficient for YIG $\alpha\approx10^{-5}$ we obtain single-photon cooperativities up to $\mathcal{C}_{0}=4G^{2}/(\kappa\alpha\omega)\sim10^{-7}$, where $\kappa\sim0.1{\rm THz}$ (from COMSOL) and $\omega$ frequency of the respective magnon mode. For a maximum allowed photon density of $10^{5}\SI{}{\micro\meter}^{-3}$, $\mathcal{C}=n_{ph}\mathcal{C}_{0}\sim10^{-2}$, a five orders of magnitude improvement with respect to the current state of the art.

\emph{Conclusion.\textendash{} }We developed a numerical method based in micromagnetics and finite-element simulations for cavity optomagnonicss with magnetic textures. We studied a microdisk where the magnetic static background is a vortex. The system presents two qualitatively distinct regimes. For thin disks the problem allows for an approximate analytical treatment, which we use to benchmark our results. For this case, we propose a heterostructure where the optical cavity surrounds the microdisk for better confinement of the optical modes. A simpler structure from the experimental point of view could be instead an optical cavity \emph{on top} of the microdisk, where the coupling is evanescent. This could provide the freedom of designing optical modes independently of the magnetic structure. For thick disks, the microdisk serves also as the optical cavity. This system presents a rich magnetic structure, and large values of optomagnonic coupling and cooperativities are in principle achievable. Coupling to other spin wave modes in microdisks, of the WGM kind~\cite{schultheiss}, could boost these values even further. The predicted values imply a significant improvement with respect to the state of the art, and are attainable within current technology. Our results pave the way for optomagnonics with magnetic textures~\cite{zueco2018,stamps2018}, including optically induced non-linear vortex dynamics (e.g. self-oscillations of the gyrotropic mode), optically mediated synchronization in vortex arrays, and exotic quantum states entangling vortex and optical degrees of freedom. Finally, our results indicate the potential of these systems for cavity-enhanced Brillouin scattering microscopy to study vortices or other magnetic structures.

\emph{Acknowledgments.\textendash{}} We thank A. Aiello for discussions and K. Usami for useful comments on the manuscript. F.M. acknowledges support through the European FET proactive network "Hybrid Optomechanical Technologies". S.V.K. acknowledges support from the Max Planck Gesellschaft through an Independent Max Planck Research Group.

\bibliography{References}
%################################################################
\newpage 
\thispagestyle{empty}
\quad 
\newpage

\maketitle
\onecolumngrid
\appendix

\renewcommand{\thefigure}{A.\arabic{figure}}
\setcounter{figure}{0} 

\section{\label{sec:Gyrotropic-Mode}Gyrotropic Mode}

In this section we calculate the local change in the magnetization $\delta\mathbf{m}(\vec{\rho,}t)$ corresponding to small oscillations of the spin with respect to the local equilibrium due to the gyrotropic mode. The obtained results are valid for the thin disk approximation. 

We can parametrize the magnetization outside of the vortex core as
\begin{equation}
\mathbf{m}(\vec{\rho})=\mathcal{C}\mathbf{e}_{\varphi}\,,
\end{equation}
with $\vec{\rho}=(\rho,\varphi)$ the polar coordinates in the system with origin at the center of the core. We describe the magnetization profile inside of the core with the ''rigid vortex'' model~\cite{guslienko_magnetic_2008-1,usov_magnetization_1993} using the following parametrization 
\begin{equation}
\mathbf{m}(\mathbf{r})=\frac{1}{\rho^{2}+b^{2}}\left(\begin{array}{c}
-2\mathcal{C}by\\
2\mathcal{C}bx\\
\mathcal{P}\left(b^{2}-\rho^{2}\right)
\end{array}\right)\,.
\end{equation}
The radius of the vortex core $b$ can be obtained approximately by energy considerations~\cite{usov_magnetization_1993}. For a disk with micrometer radius, $b$ is of the order of a few $l_{{\rm ex}}$. The time-dependent magnetization as long as the gyrotropic mode is excited can be approximated as 
\[
\mathbf{m_{{\rm ex}}}(\mathbf{r},t)=\mathbf{m}(\mathbf{r}-r_{c}(t))\approx\mathbf{m}(\mathbf{r})-\left(r_{c}(t)\cdot\nabla\right)\mathbf{m}(\mathbf{r})
\]
with $\mathbf{r}_{c}(t)=r_{c}\left[\cos(\omega_{g}t)\mathbf{e_{x}}+\mathcal{P}\sin(\omega_{g}t)\mathbf{e}_{y}\right]$, where we have ignored damping of the mode. Using the complex representation of $r_{c}(t)$ we obtain 
\begin{equation}
\delta\mathbf{m}(\mathbf{r},t)=-\frac{r_{c}}{2}\left(\frac{\partial\mathbf{m}}{\partial x}-i\mathcal{P}\frac{\partial\mathbf{m}}{\partial y}\right)e^{i\omega_{g}t}+c.c.\,.\label{eq:delta_m_g}
\end{equation}
We first consider the perturbation outside the vortex core such that $\rho\ge b$ and $\mathbf{m}_{o}(\mathbf{r})=\mathcal{C}\mathbf{e}_{\varphi}$. In a cartesian coordinate system with center at the unexcited vortex $\mathbf{r}_{0}$ we get 
\begin{equation}
\mathbf{m}_{{\rm o}}(\mathbf{r})=\frac{\mathcal{C}}{\sqrt{x^{2}+y^{2}}}\big(\begin{array}{c}
-y\\
x
\end{array}\big)\label{eq:m_vortex}
\end{equation}
with $x=\rho\cos(\varphi)$, $y=\rho\sin(\varphi)$. Hence the following expressions hold 
\begin{align}
\frac{\partial\mathbf{m}_{{\rm o}}}{\partial x} & =\mathcal{C}\frac{\sin(\varphi)}{\rho}\mathbf{e}_{\rho}\label{eq:dxdym},\\
\frac{\partial\mathbf{m}_{{\rm o}}}{\partial y} & =-\mathcal{C}\frac{\cos(\varphi)}{\rho}\mathbf{e_{\rho}}\,.\nonumber 
\end{align}
Inserting Eqs.~\eqref{eq:dxdym} into Eq.~\eqref{eq:delta_m_g} and using $e^{-i\mathcal{P}\varphi}=\cos(\varphi)-i\mathcal{P}\sin(\varphi)$ we obtain 
\begin{equation}
\delta\mathbf{m}_{{\rm o}}(\mathbf{r},t)=-\mathcal{C}\mathcal{P}\frac{r_{c}}{2\rho}e^{-i\left(\mathcal{P}\varphi-\pi/2\right)}e^{i\omega_{g}t}\mathbf{e}_{\rho}+c.c.\,,\label{eq:delta_m_out}
\end{equation}
with $\mathbf{e}_{\rho}=\left(\cos(\varphi),\sin(\varphi)\right)$. 

Inside the vortex core, the magnetization acquires an out-of-plane component. In the rigid vortex model, the vortex moves without deforming and for $\rho\le b$ it is parametrized as 
\begin{equation}
\mathbf{m}_{{\rm i}}(\mathbf{r})=\frac{1}{\rho^{2}+b^{2}}\left(\begin{array}{c}
-2\mathcal{C}by\\
2\mathcal{C}bx\\
\mathcal{P}\left(b^{2}-\rho^{2}\right)
\end{array}\right)\,.
\end{equation}
For the gyrotropic mode, using \eqref{eq:delta_m_g}, we obtain
\begin{align}
\delta m_{{\rm i}x} & =-2r_{c}\mathcal{C}\,\frac{b}{\left(\rho^{2}+b^{2}\right)^{2}}\left[2xy+i\mathcal{P}\left(\rho^{2}+b^{2}-2y^{2}\right)\right]\label{eq:delta_m_i},\\
\delta m_{{\rm i}y} & =-2r_{c}\mathcal{C}\,\frac{b}{\left(\rho^{2}+b^{2}\right)^{2}}\left[\left(\rho^{2}+b^{2}-2x^{2}\right)+i\mathcal{P}2xy\right]\nonumber, \\
\delta m_{{\rm i}z} & =4r_{c}\mathcal{P}\,\frac{\rho b^{2}}{\left(\rho^{2}+b^{2}\right)^{2}}e^{i\mathcal{P}\varphi}\:.\nonumber 
\end{align}

\begin{figure}[t!]
\includegraphics[height=0.42\textheight]{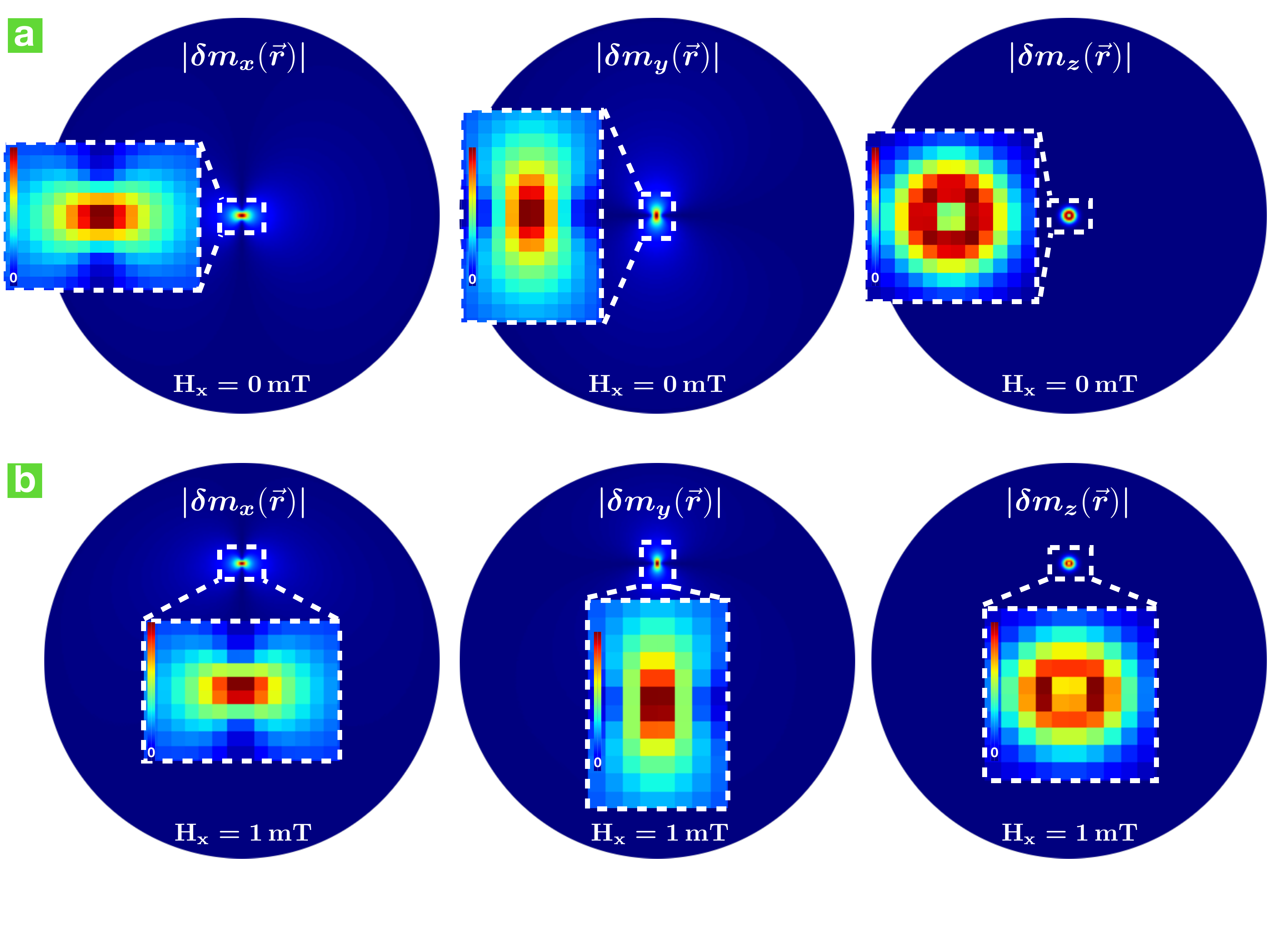} \caption{Spatial profile of the gyrotropic mode obtained by micromagnetic simulations for the thin disk ($R=\SI{1}{\micro\meter}$, $h=\SI{20}{\nano\meter}$)
for (a) $H_{x}=0$ and (b) $H_{x}=\SI{1}{\milli\tesla}$.}
\label{AFig1} 
\end{figure}

\begin{figure}[t!]
\includegraphics[width=0.7\columnwidth]{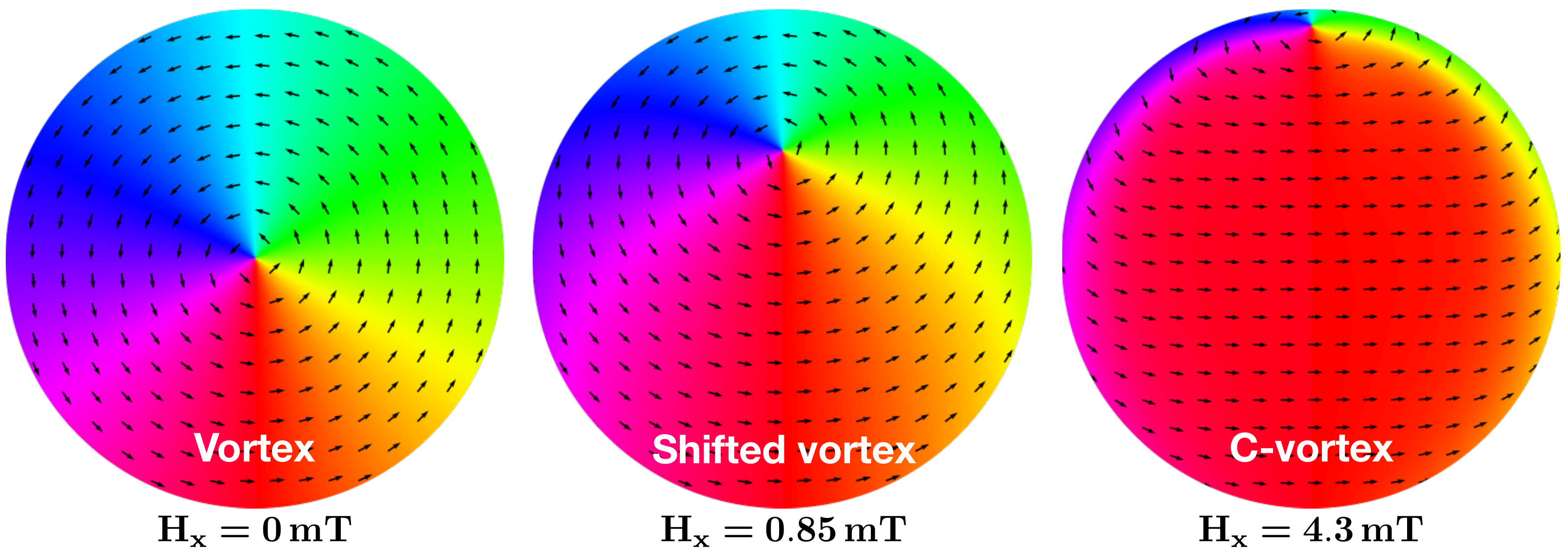} \caption{Equilibrium magnetization for increasing applied magnetic field along
the $x$ direction. Results for the thin disk ($R=\SI{1}{\micro\meter}$, $h=\SI{20}{\nano\meter}$).}
\label{AFig3} 
\end{figure}

If an external in-plane field $H_{x}$ is applied, the vortex is displaced from the center of the disk and it is deformed into a C-vortex for fields larger than a certain field $H_{x}^{n}$, see Fig. \eqref{AFig3}. The calculated mode profile in Eqs.~\ref{eq:delta_m_out} and \ref{eq:delta_m_i} are valid as long as the vortex core is not deformed, that is, for $H_{x}<H_{x}^{n}$ . Fig. \eqref{AFig1} shows the gyrotropic mode profile obtained by micromagnetics for a field $H_{x}<H_{x}^{n}$. The results are in agreement with the mode profile obtained in Eqs.~\eqref{eq:delta_m_out} and \eqref{eq:delta_m_i}.
%-----------------------------------------------------------------------------------------------------------------------------------------------------------------------
%-----------------------------------------------------------------------------------------------------------------------------------------------------------------------

\section{\label{sec:Normalization-of-the}Normalization of the Gyrotropic
Mode}

We can relate the amplitude $r_{c}$ of the gyrotropic mode to the average number of excited magnons in the mode. For this we recall that the local magnetization can be written as~\cite{hanskyrmions2017}
\begin{equation}
\mathbf{m}_{{\rm ex}}(\mathbf{r})=\mathbf{e}_{3}\sqrt{1-\frac{2g\mu_{{\rm B}}}{M_{{\rm s}}}|\psi(\mathbf{r})|^{2}}+\sqrt{\frac{g\mu_{{\rm B}}}{M_{{\rm s}}}}\left[\psi(\mathbf{r})\mathbf{e}_{+}+\psi^{*}(\mathbf{r})\mathbf{e}_{-}\right]
\end{equation}
where $\mathbf{e}_{3}=\mathbf{m}(\mathbf{r})$ corresponds to the local equilibrium direction, $\left(\mathbf{e}_{1},\mathbf{e}_{2},\mathbf{e}_{3}\right)$ is a local orthonormal basis, and $\mathbf{e}_{\pm}=\left(\mathbf{e}_{1}\pm i\mathbf{e}_{2}\right)/\sqrt{2}$. $\psi(\mathbf{r})$ is the (complex) amplitude of the spin wave, and $|\psi(\mathbf{r})|^{2}$ gives us the probability density of magnonic excitations. We can therefore identify 
\begin{equation}
\delta\mathbf{m}(\mathbf{r})=\sqrt{\frac{g\mu_{{\rm B}}}{M_{{\rm s}}}}\left[\psi(\mathbf{r})\mathbf{e}_{+}+\psi^{*}(\mathbf{r})\mathbf{e}_{-}\right]
\end{equation}
and obtain $\psi_{{\rm i}}(\mathbf{r})$ , $\psi_{{\rm o}}(\mathbf{r})$ from the corresponding expressions in Eqs.~\eqref{eq:delta_m_i} and \eqref{eq:delta_m_out}.

We proceed with finding $\psi_{{\rm i}}$. While $\mathbf{e}_{3}$ is determined by the local equilibrium magnetization, $\mathbf{e}_{\pm}$ can be chosen. Writing $\mathbf{m}(\mathbf{r})$ in cylindrical coordinates results in a natural way to chose $\mathbf{e}_{\pm}$ (in what follows we take for notational simplicity $\mathcal{C}=\mathcal{P}=1$ ),
\begin{equation}
\mathbf{e}_{3}^{{\rm i}}=\mathbf{m_{{\rm i}}}(\mathbf{r})=\frac{2b\rho}{\rho^{2}+b^{2}}\mathbf{e}_{\varphi}+\frac{b^{2}-\rho^{2}}{\rho^{2}+b^{2}}\mathbf{e}_{z}=\cos(\phi)\mathbf{e}_{\varphi}+\sin(\phi)\mathbf{e}_{z},
\end{equation}
where the second equality defines the angle $\phi$ in the plane spanned by $\mathbf{e}_{\varphi}$ and $\mathbf{e}_{z}$. If we define $\mathbf{e}_{2}^{{\rm i}}$ to be in the same plane, i.e. $\mathbf{e}_{2}^{{\rm i}}=\sin(\phi)\mathbf{e}_{\varphi}-\cos(\phi)\mathbf{e}_{z}$, then the local triad is completely determined by $\mathbf{e}_{1}^{{\rm i}}=\mathbf{e}_{2}^{{\rm i}}\times\mathbf{e}_{3}^{{\rm i}}=\mathbf{e}_{\rho}$. Hence we have $\delta\mathbf{m_{{\rm i}}}(\mathbf{r})\cdot\mathbf{e}_{-}^{{\rm i}}=0$, $\delta\mathbf{m_{{\rm i}}}^{*}(\mathbf{r})\cdot\mathbf{e}_{+}^{{\rm i}}=0$
and we can write 
\begin{align}
\psi_{{\rm i}}(\mathbf{r}) & =\sqrt{\frac{M_{{\rm s}}}{g\mu_{{\rm B}}}}\frac{\delta\mathbf{m_{{\rm i}}}(\mathbf{r})}{2}\cdot\mathbf{e}_{+}^{{\rm i}}=-\sqrt{\frac{M_{{\rm s}}}{g\mu_{{\rm B}}}}\frac{\sqrt{2}br_{c}}{b^{2}+\rho^{2}}\left[\sin(\varphi)+i\cos(\varphi)\right],\label{Dummy}\\
\psi_{{\rm i}}^{*}(\mathbf{r}) & =\sqrt{\frac{M_{{\rm s}}}{g\mu_{{\rm B}}}}\frac{\delta\mathbf{m_{{\rm i}}}^{*}(\mathbf{r})}{2}\cdot\mathbf{e}_{-}^{{\rm i}}.
\end{align}
%-----------------------------------------------------------------------------------------------------------------------------------------------------------------------
%-----------------------------------------------------------------------------------------------------------------------------------------------------------------------
Using this we obtain 
\begin{equation}
\int{\rm d^{3}}\mathbf{r}|\psi_{{\rm i}}(\mathbf{r})|^{2}=2\pi h\frac{M_{{\rm s}}}{g\mu_{{\rm B}}}\int_{0}^{b}\rho{\rm d}\rho\left(\frac{\sqrt{2}br_{c}}{b^{2}+\rho^{2}}\right)^{2}=\pi h\frac{M_{{\rm s}}}{g\mu_{{\rm B}}}r_{c}^{2}\,.
\end{equation}
Outside of the vortex core, $\mathbf{e}_{3}^{{\rm o}}=\mathbf{m}_{{\rm o}}(\mathbf{r})=\mathbf{e}_{\varphi}$ holds and the fluctuations $\delta\mathbf{m}_{{\rm o}}$ are in the plane and along $\mathbf{e}_{\rho}$ (see Eq.~\eqref{eq:delta_m_out}). Hence we simply get $\delta\mathbf{m}_{{\rm o}}/2=\psi_{{\rm o}}(\mathbf{r})\mathbf{e}_{\rho}$ and hence 
\begin{equation}
\psi_{{\rm o}}(\mathbf{r})=-i \sqrt{\frac{M_{{\rm s}}}{g\mu_{{\rm B}}}}\frac{r_{c}}{2\rho}e^{-i\varphi}\,.
\end{equation}
In this case we obtain 
\begin{equation}
\int{\rm d^{3}}\mathbf{r}|\psi_{{\rm o}}(\mathbf{r})|^{2}=\frac{\pi h}{2}\frac{M_{{\rm s}}}{g\mu_{{\rm B}}}r_{c}^{2}\int_{b}^{R}\frac{{\rm d}\rho}{\rho}=\frac{\pi h}{2}\frac{M_{{\rm s}}}{g\mu_{{\rm B}}}r_{c}^{2}\ln\left(\frac{R}{b}\right)\,.
\end{equation}
By normalizing to one magnon in the disk volume, we obtain 
\begin{equation}
r_{c}^{2}=\frac{g\mu_{{\rm B}}}{M_{{\rm s}}}\frac{1}{\pi h}\left[1+\frac{1}{2}\ln\left(\frac{R}{b}\right)\right]^{-1}\,.\label{eq:rc_normalization}
\end{equation}
For a typical microdisk with $R\approx\SI{1}{\micro\metre}$ and $b\approx\SI{10}{\nano\metre}$ we get 
\begin{equation}
r_{c}^{2}\approx\frac{1}{3}\frac{g\mu_{{\rm B}}}{M_{{\rm s}}}\frac{1}{\pi h}.\label{equ:rc}
\end{equation}
In the case of a YIG disk with $h\approx\SI{25}{\nano\metre}$ \ref{equ:rc} evaluates to $r_{c}\approx\SI{0.2}{\angstrom}$.

\section{\label{sec:Whispering-Gallery-Modes}Whispering Gallery Modes in Cylindrical Geometry}

In the limit of an infinite cylinder, Maxwell equations can be solved analytically due to the translation invariance along the cylinder axis $z$. Hence the problem can be considered as two dimensional. In the following we sketch the solution for completeness~\cite{heebnerjohnoptical2008}. 

The functional form of the electric $\psi=E_{z}$ or the magnetic $\psi=B_{z}$ field component is given by solving the Helmholtz equation
\begin{equation}
\left(\nabla^{2}+n^{2}k^{2}\right)\psi=0,
\end{equation}
respectively for the TM ($\mathbf{B}\perp\mathbf{e}_{z}$) and the TE ($\mathbf{E}\perp\mathbf{e}_{z}$) mode. In cylindrical coordinates $(r,\theta)$ with origin at the center of the disk, one obtains in the case $r<R$ 
\begin{equation}
\psi(r,\theta)=A_{m}J_{m}(kr)e^{i(\pm m\theta)}\,,
\end{equation}
with $J_{m}$ the Bessel function of the first kind. Since the magnetic mode is confined to the disk, we can focus only the solution for $r<R$. The outer solution $r>R$ however is needed to obtain the allowed values of $k$. This is given by the condition 
\begin{equation}
K\partial_{r}J{}_{m}(nkR)/J_{m}(nkR)=\partial_{r}H^{(1)}{}_{m}(kR)/H_{m}^{(1)}(kR)\label{eq:k_mp}
\end{equation}
with $H_{m}^{(1)}$ the Hankel function of the first kind and $K=n$ for a TM and $K=1/n$ for a TE mode. The solutions to this transcendental equation are of the form 
\begin{equation}
\tilde{k}_{mp}=k_{mp}-ik_{mp}''
\end{equation}
where the real part $k_{mp}$ determines the position of the resonance and the imaginary part $k_{mp}''>0$ the leaking of the mode out of the cavity, and therefore its lifetime. Additionally, $2m$ gives the number of interference maxima of the E/B-field in the azimuthal direction and $p$ the number of interference regions in the radial direction. In the following we are interested in the $\mathbf{E}$ field, which is the field relevant for the optomagnonic coupling. For the TM mode one obtains simply 
\begin{equation}
\mathbf{E}_{mp}^{{\rm TM}}=\psi_{mp}(r,\theta)\mathbf{e}_{z}\label{eq:TM mode}
\end{equation}
while for the TE mode 
\begin{equation}
\mathbf{E}_{mp}^{{\rm TE}}=\frac{i}{\varepsilon\tilde{\omega}_{mp}}\left(\frac{1}{r}\partial_{\theta}\psi_{mp}\mathbf{e}_{r}-\partial_{r}\psi_{mp}\mathbf{e}_{\theta}\right)\label{eq:TE mode}
\end{equation}
holds, where $\tilde{\omega}=\frac{c}{n}\,\tilde{k}=\omega+\frac{i\kappa}{2}$ and the subscripts $mp$ indicates that the expressions are evaluated for a particular solution $\tilde{k}=\tilde{k}_{mp}$ of Eq.~\eqref{eq:k_mp}. Since WGMs in a classical sense are located at the rim of the disk, we assume $p=1$ and hence omit this index in the following discussions. By considering well defined WGMs, $k_{m}$, determined by the boundary conditions at the rim, can be taken as real in a first approximation.

The normalization coefficient to one photon in average in the optical cavity can be found from 
\begin{equation}
\frac{\varepsilon_{0}\varepsilon}{2}\int_{V}{\rm d}\mathbf{r}|\mathbf{E}(\mathbf{r})|^{2}=\frac{\hbar\omega_{m}}{2}\,.\label{eq:NormAm}
\end{equation}
 For the TE mode we obtain 
\begin{equation}
\frac{\varepsilon_{0}\varepsilon h}{(\varepsilon\omega_{m})^{2}}|A_{m}|^{2}=\frac{\hbar\omega_{m}}{2\pi\mathcal{N}_{J}}\label{eq:NormAmExpl}
\end{equation}
 with 
\begin{equation}
\mathcal{N}_{J}=\int_{0}^{1}u{\rm d}u\left[\frac{m^{2}}{u^{2}}|J_{m}(k_{m}Ru)|^{2}+|\partial_{u}J_{m}(k_{m}Ru)|^{2}\right]\,.\label{eq:NJ}
\end{equation}

\section{\label{sec:Optomagnonic-coupling-Vortex-WGM}Optomagnonic coupling for the thin Disk}

\begin{figure}
\begin{centering}
\includegraphics[width=0.7\textwidth]{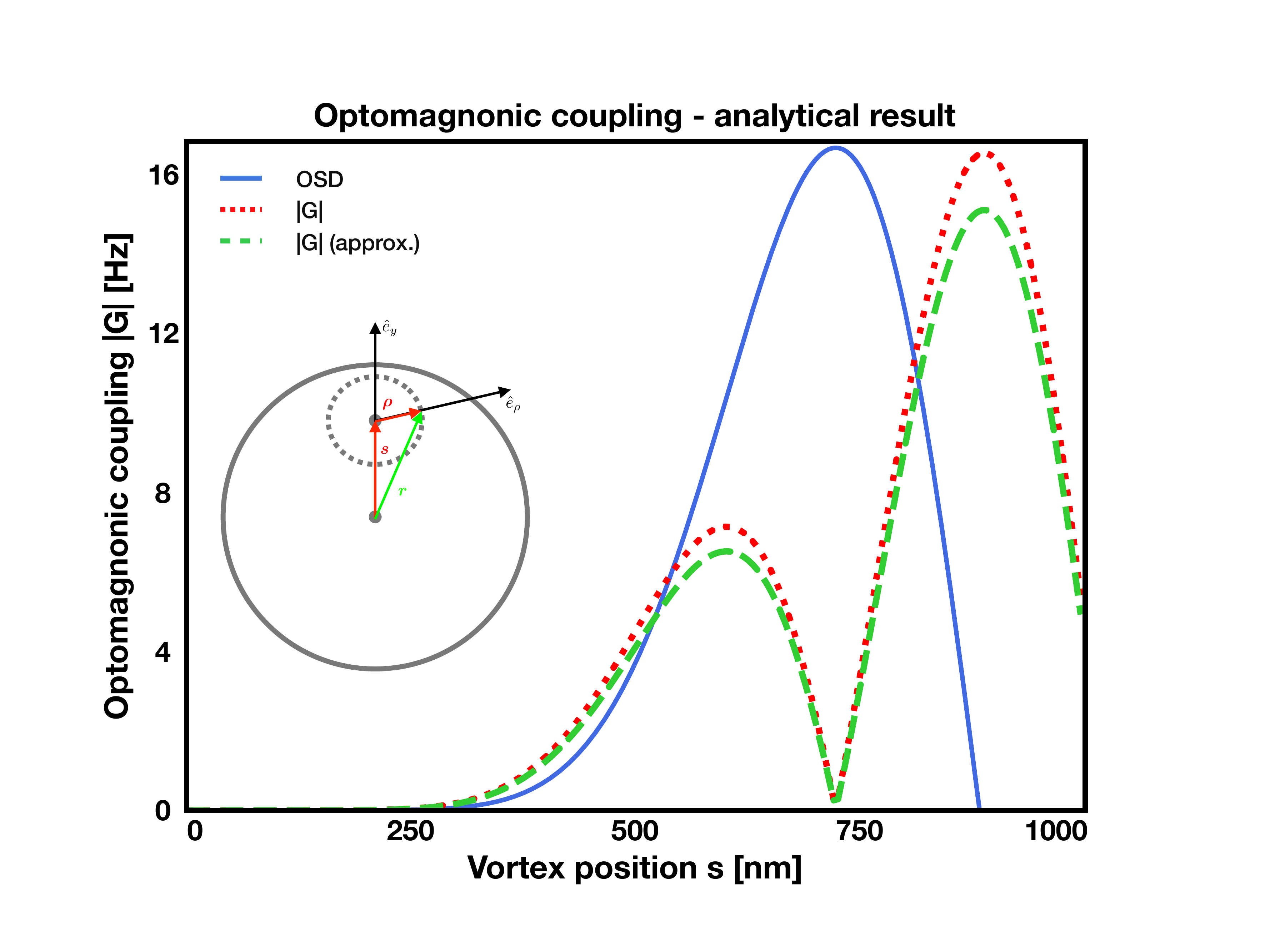}
\par\end{centering}
\caption{Red: Optomagnonic coupling for the gyrotropic mode as a function of position of the vortex for a YIG thin disk ($R=\SI{1}{\micro\meter}$), according to the analytical expression in Eq.~\eqref{eq:Gtheo}. We took a WGM with $m=6$, $\omega_{{\rm opt}}/2\pi\approx180\SI{}{\tera\hertz}$. Green: Approximated optomagnonic coupling by using a Taylor expansion of the OSD. Blue: Magnitude of the optical spin density evaluated at the vortex position (in arbitrary units). Inset: Used coordinate system.}
 \label{Fig:Gtheo}
\end{figure}

We consider an applied external magnetic field $H_{x}$ such that the vortex is displaced a distance $s$ from the disk's center, and calculate the coupling of the gyrotropic mode discussed in Sec. ~\ref{sec:Gyrotropic-Mode} to a TE-WGM as given in Eq.~\eqref{eq:TE mode}. The optical spin density vector is perpendicular to the disk plane ($z$-axis) and therefore couples only to $\delta m_{{\rm i}z}$ given in Eq.~\eqref{eq:delta_m_i}, which is finite only for $\rho\le b$. The optomagnonic coupling in this case reads
\begin{equation}
G_{m} = -i\frac{\theta_{{\rm F}}\lambda_{n}}{4\pi}\frac{\varepsilon_{0}\varepsilon}{2}h\!\int_{0}^{b}\!\!\rho~ d\rho\int_{0}^{2\pi}\!\!d\varphi~ m_{z}(\rho,\varphi) \left(\mathbf{E}_{m}^{{\rm TE}*}\times\mathbf{E}_{m}^{{\rm TE}}\right)\cdot  \mathbf{e}_{z}.
\end{equation}
Here, $(\rho,\varphi)$ are polar coordinates in the system with origin at the center of the vortex. From Eq.~\eqref{eq:TE mode} we obtain ($\tilde{r}=k_{m}r$, and $\omega_{m}=\frac{c}{n}k_{m}$) 
\begin{align}
\mathbf{E}_{\pm m}^{{\rm TE}*}\times\mathbf{E}_{\pm m}^{{\rm TE}}& =\pm i\mathbf{e}_{z}\frac{|A_{m}|^{2}}{\varepsilon^{2}\omega_{m}^{2}}\frac{k_{m}m}{r}~\partial_{\tilde{r}}|J_{m}(\tilde{r})|^{2}\,,\label{eq:ETExETE}
\end{align}
where $(\pm) m$ corresponds to $e^{\pm im\theta}$ and $\mathbf{r}=s\mathbf{e}_{y} \pm \rho\mathbf{e}_{\rho}$ (see inset of Fig.~\ref{Fig:Gtheo} for details). 
Using also Eq.~\eqref{eq:NJ} yields
\begin{equation}
G_{\pm m} = (\pm)\, 2 \pi \hbar \frac{m}{\mathcal{N}_J} \frac{\theta_{{\rm F}}\lambda_{n}}{2\pi} r_c \frac{n}{c} \frac{\omega_m^2}{4 \pi^2} \int_{0}^{b}d\rho \int_{0}^{2\pi} d\varphi ~ e^{i\varphi}\frac{\rho^{2} b^2}{\left(\rho^{2}+b^2\right)^{2}}\frac{\partial_{\tilde{r}}|J_{m}(\tilde{r})|^{2}}{r}.
\end{equation}
We note that the last factor in the integrand has a non-trivial dependence on $\varphi$, since the vortex is displaced from the center of the disk. The coefficient $r_{c}$ is determined by the number of magnons in the excited mode and given by $r_{c}\approx\SI{0.2}{\angstrom}$ for a single magnonic excitation (see Sec. \ref{sec:Normalization-of-the}). Taking a WGM with $f\approx\SI{180}{\tera\hertz}$ captured in a YIG disk with $\theta_{{\rm F}}\lambda_{n}/2\pi\approx4\times10^{-5}$ and $n \approx \sqrt{5}$, we obtain 
\begin{equation}
\frac{G_{\pm m}\!}{2\pi\hbar} = (\pm) \frac{m}{\mathcal{N}_J} \int_{0}^{b}d\rho \int_{0}^{2\pi} d\varphi ~ e^{i\varphi}\frac{\rho^{2} b^2}{\left(\rho^{2}+b^2\right)^{2}}\frac{\partial_{\tilde{r}}|J_{m}(\tilde{r})|^{2}}{r} \times (\SI{0.2}{\mega\hertz} ).
\label{eq:Gtheo}
\end{equation}
Fig.~\ref{Fig:Gtheo} shows this expression as a function of the vortex position $s$ (red doted line). As we see the result agrees reasonably well with the numerical results presented in the main text, both for the order of magnitude of the coupling ($G_{m}\approx\SI{10}{\hertz}$), and as for the non-monotonic behavior.
Additionally this plot displays the magnitude of the OSD (blue line) normalized to one and evaluated at the position of the vortex. 

As discussed in the main text, also the analytical results indicate that the optomagnonic coupling vanishes at the maximum of the OSD, meaning that the magnon mode couples effectively to the gradient of the OSD, and is maximum at the points of maximum slope.
In order to verify this conjecture we define $\tilde{r} = k_m r$ and $O(\tilde{r}) = \partial_{\tilde{r}}|J_{m}(\tilde{r})|^{2}/{\tilde{r}}$ and perform a Taylor expansion of  $O(\tilde{r})$ in small $\tilde{\rho} = k_m \rho$ around the rescaled vortex position $\tilde{s} = k_m s$ up to first order in $\tilde{\rho}$. Without giving any further proof we note that an expansion only up to first order in $\tilde{\rho}$ is sufficient for our purposes since all higher order terms give a negligible contribution. The expansion of $O(\tilde{r})$ reads
\begin{equation}
\begin{split}
O(\tilde{r}) &= O(\tilde{r})|_{\tilde{r} = \tilde{s}} + \boldsymbol{\nabla} O(\tilde{r})|_{\tilde{r} = \tilde{s}}\cdot \,{\tilde{\boldsymbol{\rho}}}\\[0.15cm]
&= O(\tilde{s}) + \tilde{\rho}~ \partial_{\tilde{\rho}} O(\tilde{r})|_{\tilde{r}=\tilde{s}}\\
&= O(\tilde{s}) + \tilde{\rho}~ \cos(\varphi)\frac{\partial}{\partial \tilde{r}}~ O(\tilde{r})|_{\tilde{r}=\tilde{s}}.
\end{split}
\label{Eq:Taylor}
\end{equation}
Inserting this expansion of $O(\tilde{r})$ into Eq.~\eqref{eq:Gtheo} and performing the change of variable $\rho \rightarrow \tilde{\rho}$ yields
\begin{equation}
\frac{G_{m}^{\,\text{app}}}{2\pi\hbar} =  \frac{m}{\mathcal{N}_J}  \frac{\theta_{{\rm F}}\lambda_{n}}{2\pi} \frac{r_c c}{4 n\pi^2}  \int_{0}^{k_m b}d\tilde{\rho} \int_{0}^{2\pi} d\varphi ~ \left[e^{i\varphi}\frac{\tilde{\rho}^{2} b^2}{\left(\frac{\tilde{\rho}^2}{k_m^2}+b^2\right)^{2}} ~O(\tilde{s}) + e^{i\varphi} \cos(\varphi)\frac{\tilde{\rho}^{3} b^2}{\left(\frac{\tilde{\rho}^2}{k_m^2}+b^2\right)^{2}} ~\left|\partial_{y}O(y)\right|_{y=\tilde{s}} \right],
\end{equation}
where the absolute value $\left|\partial_{y}O(y)\right|_{y=\tilde{s}} $ indicates we are now taking the derivative along the $\hat{e}_y$ axis, and the $\pm$ has been absorbed in the definition of $m$. We see that, due to symmetry, the  the only surviving term is the one proportional to $\cos{\phi}^2$. Performing the integral we obtain %-----------------------------------------------------------------------------------------------------------------------------------------------------------------------
%-----------------------------------------------------------------------------------------------------------------------------------------------------------------------
\begin{equation}\label{Gmapprox}
\frac{G_{m}^{\,\text{app}}}{2\pi\hbar} =  \frac{m}{\mathcal{N}_J} \frac{\theta_{{\rm F}}\lambda_{n}}{2\pi} r_c k_m^3 \frac{\omega_m}{16 \pi}b^2  \left[\log(4)-1\right] ~ \left|\partial_{y}O(y)\right|_{y=\tilde{s}}\\
\end{equation}
For WGM with $f\approx\SI{180}{\tera\hertz}$ in a YIG disk with $\theta_{{\rm F}}\lambda_{n}/2\pi\approx4\times10^{-5}$ and $n \approx \sqrt{5}$ this yields
\begin{equation}
\frac{G_{\pm m}^{\,\text{app}}}{2\pi\hbar} = (\pm) \frac{m}{\mathcal{N}_J} ~\left|\partial_{y}O(y)\right|_{y=\tilde{s}} \times (\SI{400}{\hertz}).
\label{Eq:GApprox}
\end{equation}
This expression is also shown in Fig.~\ref{Fig:Gtheo} (green line). As we see the approximate expression of the coupling reproduces very well the exact coupling given in Eq.~\eqref{eq:Gtheo}, up to a small multiplicative factor. Including higher order terms in the Taylor expansion accounts for this small difference. Since  $O(r)$ is proportional to the OSD, 
\begin{equation}
\left|\partial_{y}O(y)\right|_{y=\tilde{s}} = \frac{\mathcal{N}_J}{m} \frac{2 \pi \varepsilon_0 c^2 h}{\hbar \omega_m^3}~\left| \partial_{y} \left[(\mathbf{E}_{m}^{{\rm TE}*}\times\mathbf{E}_{m}^{{\rm TE}})(y)\cdot \hat{e}_z\right]\right|_{y=\tilde{s}} .
\end{equation}
this concludes our proof. Using this expression together with Eqs.~\eqref{eq:NormAmExpl}, \eqref{Gmapprox} and \eqref{equ:rc} we obtain Eq.~(3) in the main text. 

Note that in Fig.~3 in the main text, both the optomagnonic coupling and the OSD are plotted as a function of the external field $H_{x}$, using the nonlinear dependence between the vortex position $s$ and the external magnetic field found by micromagnetic simulations and shown in Fig.~\ref{AFig7}.  (the OSD does not depend on Hx, but we translate the vortex position into magnetic field). Due to this nonlinear dependence, the point of maximum slope of the OSD plotted as a function of the magnetic field does not coincide with that one as a function of position (in particular, the steepness of the slope right and left from the maximum of the OSD is inverted, compare Fig.~\ref{AFig7} and Fig. 3). In Fig. 3 we plot the optomagnonic coupling as a function of Hx since this is the externally tunable parameter.

We end this section by noting that in Eq.~\eqref{eq:Gtheo} and the corresponding Fig.~\ref{Fig:Gtheo}, we assumed that under the application of an external field $H_{x}$, the vortex remains undeformed until the rim of the disk. This is true up to $H_{x}\approx H_{x}^{n}$. Beyond these fields, the vortex core is elongated, forming a small domain wall in the form of a C. The magnon modes in the presence of this distorted vortex differ from the ideal case~\cite{rivkin_analysis_2007,aliev_spin_2009} used in this analytical calculation. Eq.~\eqref{eq:Gtheo} therefore must be taken as an approximate expression for the optomagnonic coupling. Additionally we also have neglected the imaginary part of the wave vector for the WGM.\\

\section{\label{sec:Vortex-position-Vs.}Vortex position vs. external magnetic field}

\begin{figure}[t!]
\centering{}\includegraphics[width=0.7\columnwidth]{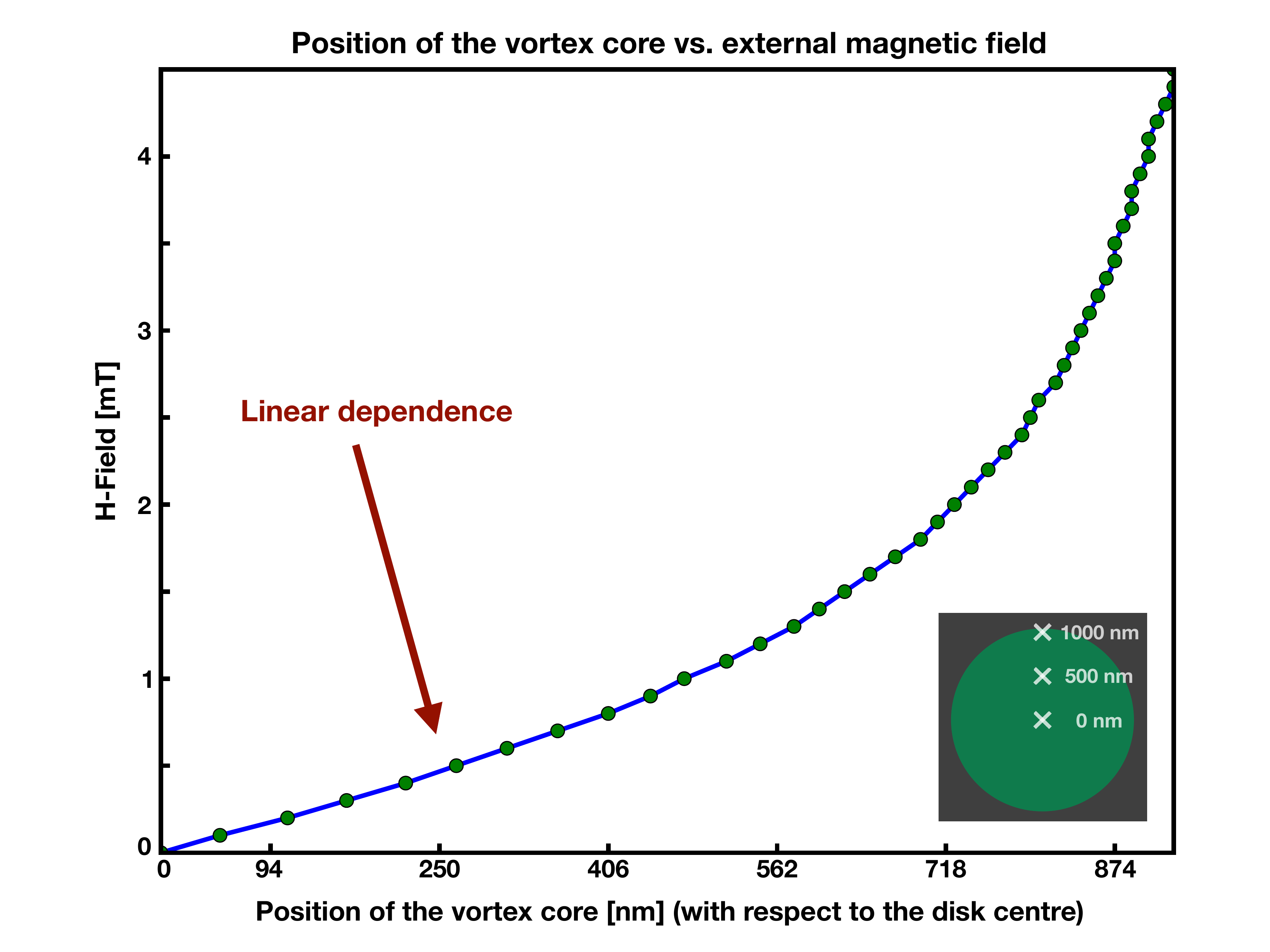}
\caption{External magnetic field as a function of the position of the vortex. Results for the thin disk ($R=\SI{1}{\micro\meter}$, $h=\SI{20}{\nano\meter}$).}
\label{AFig7}
\end{figure}

The vortex core can be shifted towards the rim by applying an external magnetic field. Hence the vortex core position can be related to the magnitude of this external field. The vortex's displacement is linear with the field for $H_{x}\lesssim H_{x}^{n}$ and one can write $s=R\chi_{{\rm m}}H_{x}$, where the magnetic susceptibility $\chi_{{\rm m}}(R,\beta)$ is defined via $\left\langle \mathbf{m}\right\rangle _{{\rm V}}=\chi_{{\rm m}}\mathbf{H}$ and its parametric dependence is given by $\chi_{{\rm m}}(R,\beta)$~\cite{guslienko_magnetization_2001}. For higher magnetic field however this dependence deviates from linear. We obtained the position of the vortex as a function of magnetic field using MuMax$^{3}$. The results are plotted in Fig.~\ref{AFig7}. 
We used these results to relate position and magnetic field in Fig.~4 %\ref{Fig3} 
of the main text.
%-----------------------------------------------------------------------------------------------------------------------------------------------------------------------
%-----------------------------------------------------------------------------------------------------------------------------------------------------------------------

\section{\label{Optical-simulations}Optical Simulations}

In order to obtain the optical eigenmodes of the cylindrical cavity we used the finite element simulation tool COMSOL Multiphysics. The simulated geometry consists of a magnetic disk surrounded by an air cylinder, with dimensions listed in Table \ref{Table2}. The air has to be taken into account, since the electric field captured in the disk can leak out at its boundaries. The size of the air cylinder was chosen such that the evanescent light has at least $3\cdot\lambda\approx\SI{4.5}{\micro\metre}$ before it reaches the boundaries. We work with the insulating magnetic material YIG with the following parameters 
\begin{align}
\varepsilon_{r} & =5\\
\mu_{r} & =1\\
\sigma & =0\,,
\end{align}
where $\varepsilon_{r}$ ($\mu_{r}$ ) is the relative permittivity (permeability) and $\sigma$ the conductivity. The air is simulated with $\varepsilon_{r}=\mu_{r}=1$ and $\sigma=0$. Due to the small height of the thin YIG disk the modes are very leaky causing the quality factor to be very low. In order to prevent that we confine the modes in the disk by sandwiching the YIG disk with two $\text{Si}_{3}\text{N}_{4}$ disks ($r=\SI{1}{\micro\metre},~h=\SI{800}{\nano\metre}$). The corresponding material parameters of $\text{Si}_{3}\text{N}_{4}$ are 
\begin{align}
\varepsilon_{r} & =4\\
\mu_{r} & =1\\
\sigma & =0\,.
\end{align}
%-----------------------------------------------------------------------------------------------------------------------------------------------------------------------
\begin{figure}[t!]
\includegraphics[height=0.42\paperwidth]{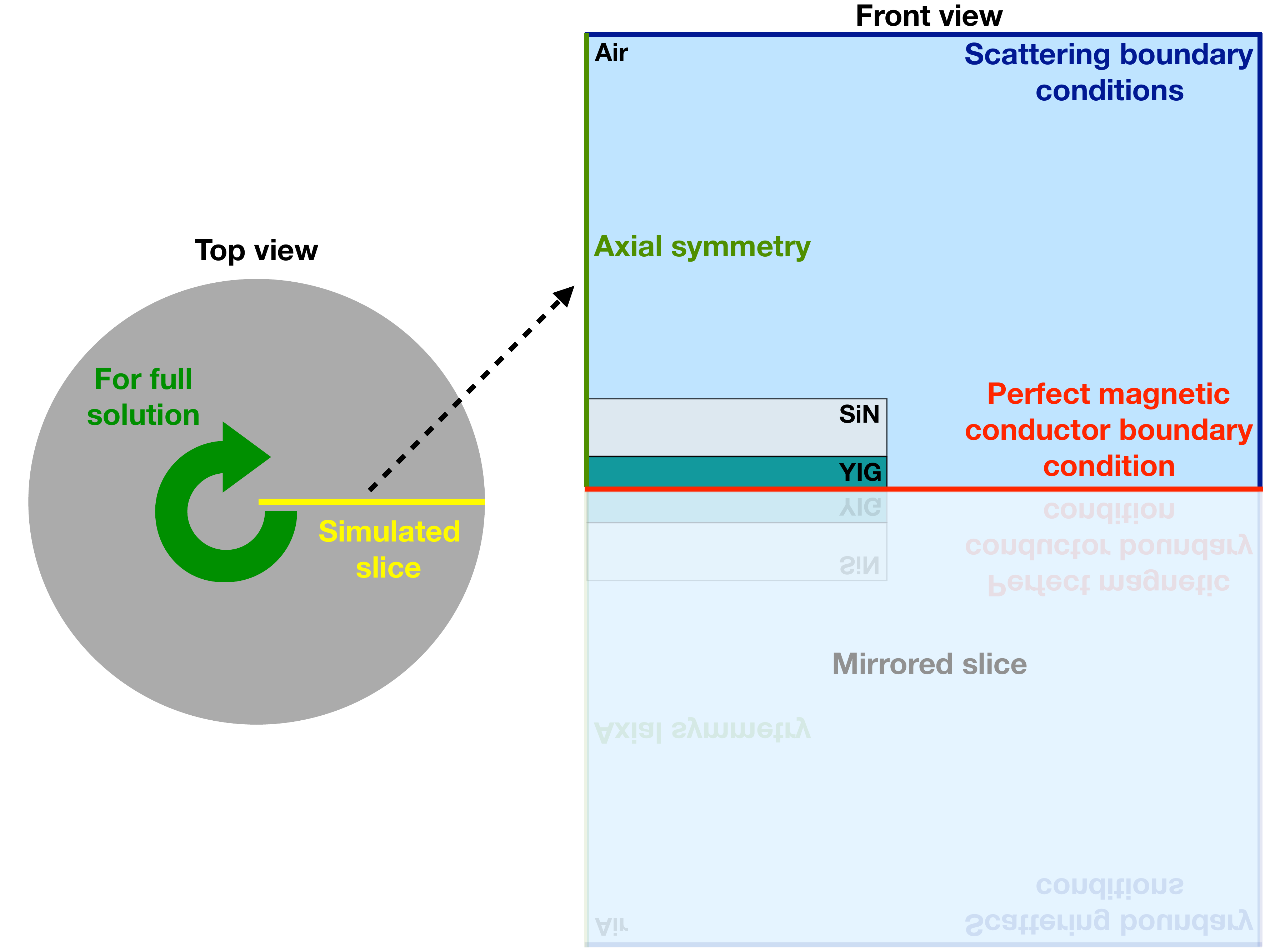} \caption{Sketch of the geometry used in the optical simulations. In the case of the thick disk, no SiN layer is present.}
\label{AFig6} 
\end{figure}
%-----------------------------------------------------------------------------------------------------------------------------------------------------------------------
\begin{table}[b!]
\begin{tabular}{|c|c|c|c|c|}
\hline 
 & $r_{\text{YIG}}$ {[}\SI{}{\micro\metre}{]}  & $h_{\text{YIG}}$ {[}\SI{}{\nano\metre}{]}  & $r_{\text{Air}}$ {[}\SI{}{\micro\metre}{]}  & $h_{\text{Air}}$ {[}\SI{}{\micro\metre}{]}\tabularnewline
\hline 
\hline 
Thin disk  & 1 & 20 & 5.5 & 5.32\tabularnewline
\hline 
Thick disk & 2 & 500 & 5.5 & 4.5\tabularnewline
\hline 
\end{tabular}
\caption{Dimensions of the YIG disk and respective air cylinder used in the simulations.}
\label{Table2}
\end{table}

\begin{table} [b!]
\begin{tabular}{|c|c|}
\hline 
Material  & Maximum element size {[}\SI{}{\nano\metre}{]}\tabularnewline
\hline 
\hline 
YIG  & 15\tabularnewline
\hline 
$\text{Si}_{3}\text{N}_{4}$  & 20\tabularnewline
\hline 
Air  & 100\tabularnewline
\hline 
\end{tabular}
\caption{Mesh element size for the different simulated system components.}
\label{Table3}
\end{table}

We use three ``Free triangular'' mesh grids, one for each material. In the YIG and $\text{Si}_{3}\text{N}_{4}$ domains we choose a finer mesh, whereas in the air domain the mesh can be increased without losing accuracy, since the optical field is concentrated closely to the disks. Details are found in table \ref{Table3}.

Due to the axial symmetry of the whole geometry we can use the ``2D Axis symmetric space dimension'' in order to save simulation time by simulating just one slice, see Fig.~\ref{AFig6}. We used the ``Electromagnetic waves, Frequency domain'' package of COMSOL's RF module which solves for
\begin{equation}
\vec{\bigtriangledown}\times\frac{1}{\mu_{r}}\left(\vec{\bigtriangledown}\times\vec{E}\right)-k_{0}^{2}\left(\varepsilon_{r}-\frac{i\sigma}{\omega\varepsilon_{0}}\right)\vec{E}=0\label{eq:HelmoltzComsol}
\end{equation}
with $k_{0}$ the vacuum wave number, $\omega$ the angular frequency and $\vec{E}$ the electric field. As the finite element method requires a finite-sized modeling domain, we need to limit the modeled air stack to a finite size. To account for leakage from the optical mode, we apply scattering boundary conditions at the surface of the air cylinder 
\[
\vec{n}\times\left(\vec{\bigtriangledown}\times\vec{E}\right)-ik\vec{n}\times\left(\vec{E}\times\vec{n}\right)=0
\]
with $\vec{n}$ the normal vector. This should avoid reflection of the electric field at the boundaries of the air cylinder, and as a consequence the obtained eigenfrequencies are complex with an imaginary part describing the loss scattered out of the the disk cavity. To find the TE modes, we simulate the upper half of the whole geometry and apply a ``perfect magnetic conductor'' boundary condition ($\vec{n}\times\vec{H}=0$) to the bottom surface of the cut geometry, see Fig.~\ref{AFig6}. The full solution is obtained by mirroring with respect to the bottom plane. Therefore within our definition, the TE (-like) modes are even under a vector-parity operation.

We search for relatively well confined optical WGMs. The micrometer scale of the system will result in low possible azimuthal numbers for the modes, and modest quality factors. We find possible candidates for $m=6$ in the case of the thin and $m=11$ in the case of the thick disk. Solving Eq.~\eqref{eq:HelmoltzComsol} using these values we obtain modes with an eigenfrequency of $218+i\SI{0.05}{\tera\hertz}$ (thin disk) and $184+i\SI{0.03}{\tera\hertz}$ (thick disk), respectively.

\section{\label{Micromagnetic-simulations}Micromagnetic Simulations}

In order to calculate the magnetics we use the GPU accelerated micromagnetic simulation tool MuMax$^{3}$~\cite{vansteenkiste_design_2014-1}. We simulate both a thin and a thick YIG disk with corresponding dimensions as given in table \ref{Table2}. The material parameters used are listed in table \ref{Table4}. 

%-----------------------------------------------------------------------------------------------------------------------------------------------------------------------
\begin{figure}[t]
\includegraphics[height=0.3\textheight]{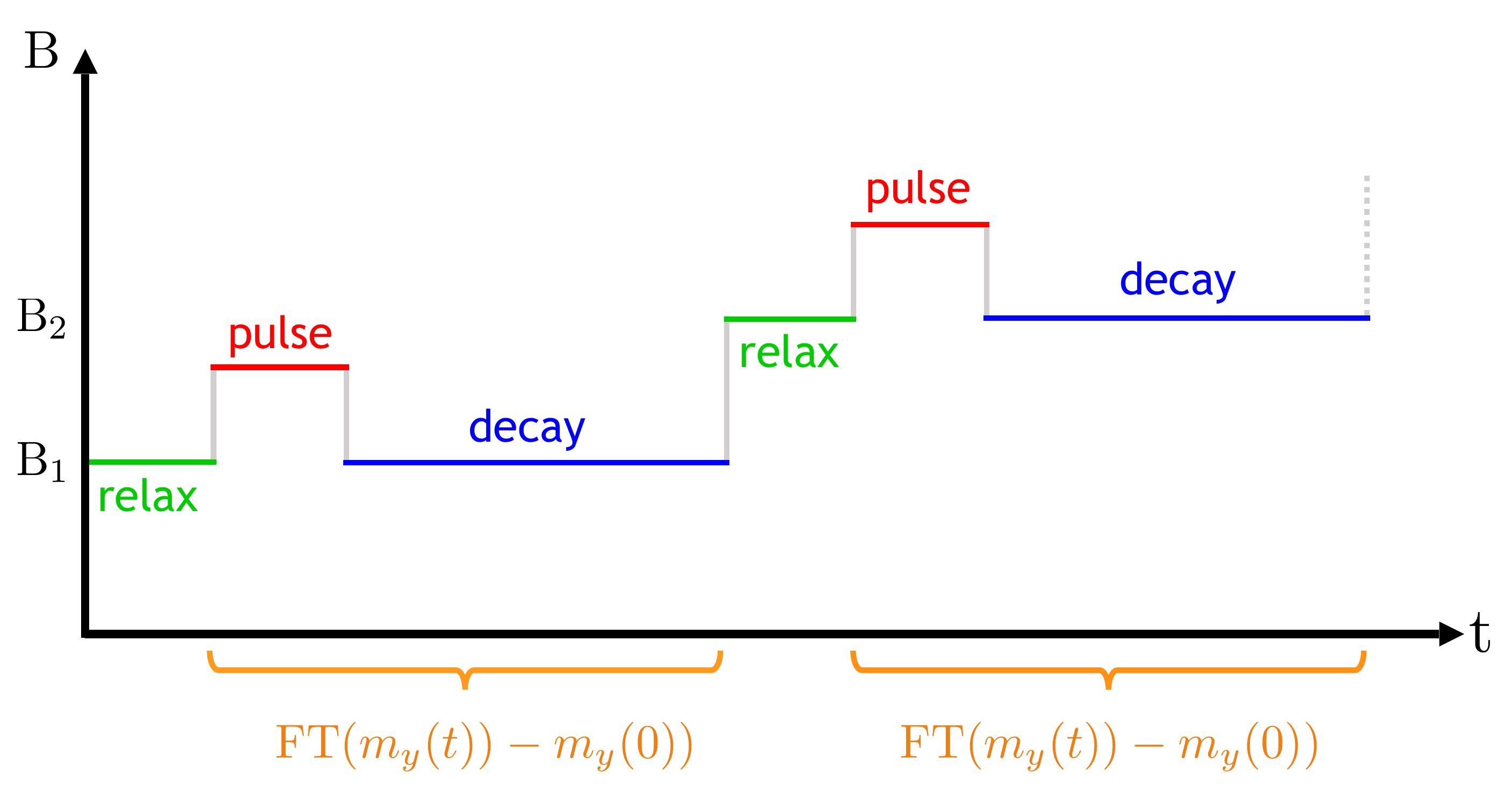}
\caption{Protocol used in the micromagnetic simulations to obtain the magnon
modes.}
\label{Fig:Protocol}
\end{figure}
%-----------------------------------------------------------------------------------------------------------------------------------------------------------------------

The two different meshes we use can be found in table \ref{Table5}, where $n_{i}$ denotes the amount of cells in the direction $i\in(x,y,z)$ and $s_{i}$ the side length. Here we pay attention that the amount of the cells is a power of two or at least has small prime factors (e.g. $2,~5$)~\cite{vansteenkiste_design_2014-1}. Furthermore the cell size should not exceed the dipole interaction length of YIG of $l_{ex}\approx\SI{12.7}{\nano\metre}$ in order to ensure that we are able to resolve the finest magnetization structures. 

\begin{table}[b!]
\begin{tabular}{|c|c|}
\hline 
Parameter  & Value \tabularnewline
\hline 
\hline 
$M_{\text{sat}}$  & $140$kA/m \tabularnewline
\hline 
$A_{\text{ex}}$  & $2$pJ/m \tabularnewline
\hline 
$K_{c1}$  & $-610$J/$\text{m}^{3}$ \tabularnewline
\hline 
anisotropy axis & $\hat{z}$ \tabularnewline
\hline 
$\alpha$  & $1$ (relax), $0.008$ (evolution) \tabularnewline
\hline 
\end{tabular}
\caption{Magnetic parameters for YIG as used in the micromagnetic simulations.}
\label{Table4}
\end{table}

\begin{table}[b!]
\begin{tabular}{|c|c|c|c|c|c|c|}
\hline 
Disk  & $n_{x}$  & $n_{y}$  & $n_{z}$  & $s_{x}~[\SI{}{\nano\metre}]$  & $s_{y}~[\SI{}{\nano\metre}]$  & $s_{z}~[\SI{}{\nano\metre}]$ \tabularnewline
\hline 
\hline 
Thin disk  & 256 & 256 & 5 & $7.8125$  & $7.8125$  & 4\tabularnewline
\hline 
Big disk  & 320 & 320 & 40 & $12.5\cdot10^{3}$  & $12.5\cdot10^{3}$  & 12.5\tabularnewline
\hline 
\end{tabular}
\caption{Mesh for micromagnetic simulations.}
\label{Table5}
\end{table}

After the geometry and the mesh are set we initialize the magnetization
with a random configuration in the case of the thin disk and already
with a vortex in the case of the big disk in order to save simulation
time. Afterwards the system is relaxed to its ground state for zero-applied
magnetic field in both cases.\\
 To find the magnon modes, we apply the following general procedure
\cite{losby_torque-mixing_2015}:
\begin{enumerate}
\item Application of a particular external magnetic field and relaxation
of the system to its corresponding ground state.
\item Excitation of the system with a short square pulse ($\SI{1.5}{\nano\second}$)
with a strength of $\SI{0.1}{\milli\tesla}$ into the $x$ direction.
\item Evaluation of the magnon mode spectrum by Fourier transforming the
time evolution of e.g. $m_{y}(t)-m_{y}(0)$. 
\end{enumerate}
In order to obtain a full mode spectrum as a function of the external
applied magnetic field, the above described procedure has to be applied
for each magnetic field separately, starting with the evolved magnetization
state of the previous B-field step. A sketch of this procedure is
found in Fig.~\ref{Fig:Protocol} 

\begin{figure}[t!]
\includegraphics[height=0.5\paperwidth]{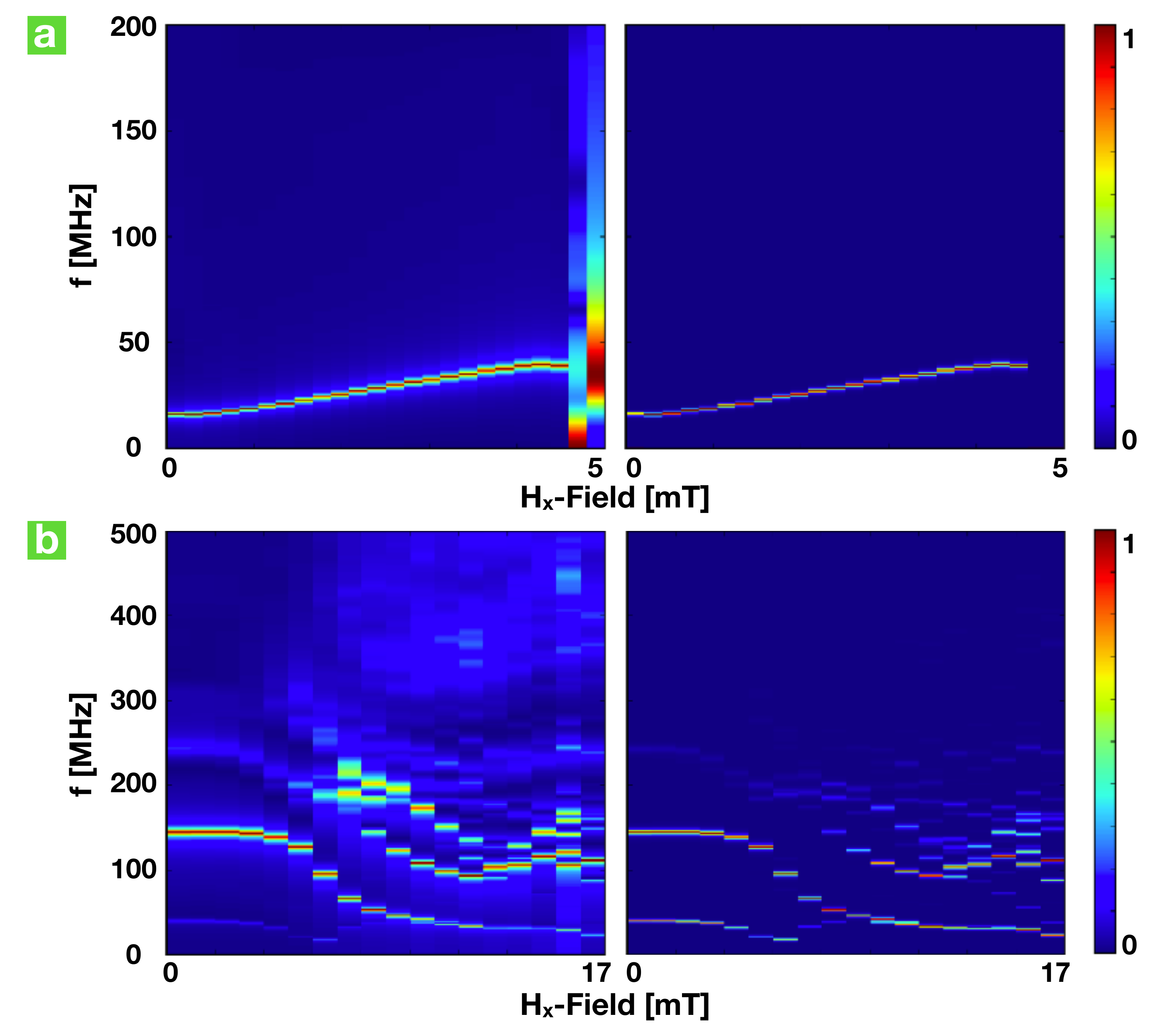}
\caption{Magnon modes spectra as a function of applied external field for the (a) thin disk ($R=\SI{1}{\micro\meter}$, $h=\SI{20}{\nano\meter}$) and (b) thick disk  ($R=\SI{2}{\micro\meter}$, $h=\SI{500}{\nano\meter}$). For the left column a short cutoff window was taken, for the right column a Hanning window.}
\label{Fig:spectra}
\end{figure}

\emph{Magnon spectrum - thin disk \textendash{}} In this case the
above approach is applied between the magnetic fields $0$ and $\SI{4.5}{\milli\tesla}$
in steps of $\SI{0.5}{\milli\tesla}$, where the system is evolved
for $\SI{2}{\micro\second}$ (-$\SI{1.5}{\nano\second}$) saving the
whole magnetization pattern each $\SI{2.5}{\nano\second}$ which gives
$2\cdot400$ time steps. With this chosen time settings we make the
frequency range $0$ to ${400}/{2}\,\SI{}{\mega\hertz}$ accessible
with a frequency resolution of $\SI{0.5}{\mega\hertz}$. The factor
of $2$ is due to symmetry of the Fourier transformation with respect
to positive and negative values. The spectrum is shown in Fig.~\ref{Fig:spectra} (a).

\emph{Magnon spectrum - thick disk \textendash{}} In this case the
excitation procedure is adopted between the external fields $0$ and
$\SI{18}{\milli\tesla}$ in steps of $\SI{1}{\milli\tesla}$, where
the system is evolved for $\SI{1}{\micro\second}$ saving the whole
magnetization each $\SI{1}{\nano\second}$ resulting in $2\cdot500$
time steps. With these settings we make the frequency range from $0$
to ${1}/{2}\,\SI{}{\giga\hertz}$ accessible with a frequency
resolution of $\SI{1}{\mega\hertz}$. The spectrum is shown in Fig.
\ref{Fig:spectra} (b). 
%-----------------------------------------------------------------------------------------------------------------------------------------------------------------------
%-----------------------------------------------------------------------------------------------------------------------------------------------------------------------

\emph{Spatial dependence of the magnon modes\textendash{}} To obtain
the spatial dependence of the magnon modes the same procedure as described
above can be applied for each particular B-field, but by using a space
dependent Fourier transformation on the magnetization components 
\begin{equation}
m_{i}(\vec{r},f)=\int dt~\left[m_{i}(\vec{r},t)-m_{i}(\vec{r},0)\right]e^{i2\pi ft},~i={x,y,z}\label{eq:FFTmag}
\end{equation}
where the Fourier transformation gets applied to the time evolution
of each simulation cell separately. 

\section{\label{sec:Normalization-of-the-1}Normalization of the coupling}

In the main text we give values of optomagnonic coupling per photon per magnon. The optical field strength obtained with COMSOL has to be scaled to have an average photon number of one in the optical cavity. The photon number is obtained by dividing the total energy of the WGM, by the energy of one photon 
\begin{equation}
N_{\text{Ph}}=\frac{E_{\text{e}}^{\text{o}}+E_{\text{m}}^{\text{o}}}{\hbar\omega_{opt}}
\end{equation}
where $E_{\text{e}}^{\text{o}}$ represents the energy of the electric field, $E_{\text{m}}^{\text{o}}$ the energy of the magnetic field and $\omega_{opt}$ the eigenfrequency optical mode. Numerically this is easily obtained by a ``Global evaluation'' in COMSOL. 

In order to calculate the magnon number we need the total energy contained in the magnon mode. This can be obtained numerically by inserting
the obtained mode profile for a given mode $\delta\mathbf{m}(\mathbf{r})$ (see Eq.~\eqref{eq:FFTmag}) back into MuMax$^{3}$ and computing the energy. Since the computed magnetization profile given by Eq.~\eqref{eq:FFTmag} is complex, we write the total energy of the mode as
\begin{align}
E_{\text{tot}}^{\text{m}} & =\frac{1}{2}\int{\rm d}\mathbf{r}\left[E^{{\rm m}}\left(\text{Re}\left\{ \delta\mathbf{m}(\mathbf{r})+\mathbf{m}_{0}(\mathbf{r})\right\} \right)+E^{{\rm m}}\left({\rm Im}\left\{ \delta\mathbf{m}(\mathbf{r})+\mathbf{m}_{0}(\mathbf{r})\right\} \right)\right]
\end{align}
where $\mathbf{m}_{0}(\mathbf{r})$ is the magnetic ground state (note that for $|\delta\mathbf{m}|/|\mathbf{m}_{0}|$ small, $E_{\text{tot}}^{\text{m}}$ is quadratic in $\delta\mathbf{m}$). The magnon number in the excited mode is obtained by 
\[
N_{\text{Mag}}=\frac{E_{\text{tot}}^{\text{m}}(\mathbf{m})-E_{\text{tot}}^{\text{m}}(\mathbf{m}_{0})}{\hbar\omega_{mag}},
\]
where $\omega_{mag}$ is the eigenfrequency of the chosen magnon mode. 

The coupling Hamiltonian is quadratic in the photon operators and linear in the magnon operators. The full normalization is therefore accomplished by dividing the computed coupling by $N_{\text{ph}}\sqrt{N_{\text{mag}}}$.

\end{document}